\input amstex
\magnification 1200
\TagsOnRight
\def\qed{\ifhmode\unskip\nobreak\fi\ifmmode\ifinner\else
 \hskip5pt\fi\fi\hbox{\hskip5pt\vrule width4pt
 height6pt depth1.5pt\hskip1pt}}
\NoBlackBoxes
\baselineskip 18.5 pt
\parskip 5 pt
\def\stretch {\noalign{\medskip}}
\define \bC {\bold C}
\define \bCp {\bold C^+}
\define \bCm {\bold C^-}
\define \bCpb {\overline{\bold C^+}}
\define \bCmb {\overline{\bold C^-}}
\define \ds {\displaystyle}
\define \bR {\bold R}
\define \bm {\bmatrix}
\define \endbm {\endbmatrix}

\centerline {\bf DARBOUX TRANSFORMATION}
\vskip -5 pt
\centerline {\bf FOR THE DISCRETE SCHR\"ODINGER EQUATION}

\vskip 7 pt
\centerline {Tuncay Aktosun}
\vskip -8 pt
\centerline {Department of Mathematics}
\vskip -8 pt
\centerline {University of Texas at Arlington}
\vskip -8 pt
\centerline {Arlington, TX 76019-0408, USA}
\vskip -8 pt
\centerline {aktosun\@uta.edu}

\vskip 10 pt
\centerline {Abdon E. Choque-Rivero}
\vskip -8 pt
\centerline {Instituto de F\'{\i}sica y Matem\'aticas}
\vskip -8 pt
\centerline {Universidad Michoacana de San Nicol\'as de Hidalgo}
\vskip -8 pt
\centerline {Ciudad Universitaria, C.P. 58048}
\vskip -8 pt
\centerline {Morelia, Michoac\'an, M\'exico}
\vskip -8 pt
\centerline {abdon\@ifm.umich.mx}

\vskip 10 pt
\centerline {Vassilis G. Papanicolaou}
\vskip -8 pt
\centerline {Department of Mathematics}
\vskip -8 pt
\centerline {National Technical University of Athens}
\vskip -8 pt
\centerline {Zografou Campus}
\vskip -8 pt
\centerline {157 80, Athens, Greece}
\vskip -8 pt
\centerline {papanico\@math.ntua.gr}

\noindent {\bf Abstract}:
The discrete Schr\"odinger equation on a half-line lattice
with the Dirichlet boundary condition is considered
when the potential is real valued, is summable, and has a finite first
moment. The Darboux transformation formulas are derived from first principles
showing how the potential and the wavefunction change when a bound state is
added to or removed from the discrete spectrum of the corresponding Schr\"odinger operator
without changing the continuous spectrum. This is done by explicitly evaluating the change in the
spectral density when a bound state is added or removed and also by determining
how the continuous part of the spectral density changes.
The theory presented
is illustrated with some explicit examples.

\vskip 5 pt
\par \noindent {\bf Mathematics Subject Classification (2010):}
39A70 47B39 81U15 34A33
\vskip -8 pt
\par\noindent {\bf Short title:} Discrete Darboux transformation
\vskip -8 pt
\par\noindent {\bf Keywords:} Discrete Schr\"odinger equation,
Darboux transformation, spectral density,
\vskip -8 pt
spectral function, Gel'fand-Levitan method

\newpage

\noindent {\bf 1. INTRODUCTION}
\vskip 3 pt

Our goal in this paper is to analyze the Darboux transformation for the discrete
Schr\"odinger equation on the half-line lattice with the Dirichlet boundary condition. In the Darboux transformation, the continuous part of the corresponding Schr\"odinger
operator is unchanged and only the discrete part of the spectrum is changed by adding or removing a finite number of
discrete eigenvalues to the spectrum. We can view the process of adding or removing
discrete eigenvalues as changing the ``unperturbed" potential and the
``unperturbed"
wavefunction into the ``perturbed" potential and the ``perturbed" wavefunction, respectively. Hence, our goal is to present the Darboux transformation formulas
  at the potential level and
  at the wavefunction level, by expressing the change
in the potential and in the wavefunction
   in terms of quantities related to the perturbation and the
   unperturbed quantities.

The Darboux transformation was termed to honor the work of French mathematician
Gaston Darboux [8], and it is useful for various reasons. For example,
it allows us to produce explicit solutions to
differential or difference equations by perturbing an already
known explicit solution. As another example, we can mention that Darboux transformations
for certain nonlinear partial differential
equations or nonlinear partial differential-difference
equations yield so-called soliton solutions, which
have important applications [14] in wave propagation of electromagnetic waves
and surface water waves. We refer the reader to the existing
literature [4,9,14-16] on the wide applications of Darboux transformation, and
in our paper we concentrate on the mathematical
aspects of the Darboux transformation
for the Schr\"odinger equation
on the half-line lattice with the Dirichlet boundary condition.

On the half-line lattice the discrete Schr\"odinger equation is given by
$$-\psi_{n+1}+2 \psi_n-\psi_{n-1}+V_n \psi_n=\lambda\, \psi_n,\qquad n\ge 1,\tag 1.1$$
where $\lambda$ is the spectral parameter, $n$ is the spacial independent variable
taking positive integer values, and
the subscripts are used to denote the dependence on $n.$
Thus, $\psi_n$ denotes the value of the wavefunction at
$n$ and $V_n$ denotes
the value of the potential at $n.$
The point $n=0$ corresponds to the boundary.
We remark that (1.1) is the analog of the half-line
Schr\"odinger equation
$$-\psi''+V(x)\,\psi=\lambda\,\psi, \qquad x>0,
\tag 1.2$$
where $\lambda$ is the spectral parameter,
the prime denotes the $x$-derivative,
$\psi$ is the wavefunction, and $V(x)$ is the potential.
The point $x=0$ corresponds to the boundary.
In analogy to (1.2), we can use (1.1) to describe [17] the behavior of a quantum
mechanical particle on a half-line lattice (such as a crystal)
experiencing the force at each lattice point $n$ resulting from the
potential $V_n.$

In order to determine the spectrum of the
corresponding Schr\"odinger operator
related to (1.1) and to identify a square-summable solution in $n$
as an eigenfunction, we must impose
a boundary condition on square-summable wavefunctions at $n=0.$ In applications
related to quantum mechanics, it is appropriate
to use the Dirichlet boundary condition at $x=0$ for (1.2), i.e.
$$\psi(0)=0,$$
and hence we impose the Dirichlet
boundary condition at $n=0$ for (1.1), i.e.
$$\psi_0=0.\tag 1.3$$
The spectrum of the corresponding
operator for (1.2) is well understood when the potential
$V(x)$ is real valued and satisfies the so-called
$L^1_1$-condition [4] given by
$$\int_0^\infty dx\,(1+x)\,|V(x)|<+\infty.\tag 1.4$$
Similarly, we assume that $V_n$ is real valued
and satisfies the analog of (1.4) given by
$$\sum_{n=1}^\infty \left(1+n\right) |V_n|<+\infty.\tag 1.5$$
Clearly, (1.5) is equivalent to
$$\sum_{n=1}^\infty n\, |V_n|<+\infty.\tag 1.6$$
The class of real-valued potentials
$V(x)$ satisfying (1.4) is usually known [4] as the Faddeev class.
Similarly, we refer to the set
of real-valued potentials $V_n$ satisfying
(1.5), or equivalently (1.6), as the Faddeev class.
The existence of the first moments in (1.4) and (1.5)
assures that the number of discrete eigenvalues
for each of the corresponding Schr\"odinger operators
is finite.

Our paper is organized as follows. In Section~2 we present the
appropriate preliminaries involving the Jost solution and
the regular solution to (1.1); the Schr\"odinger
operator, the scattering states, the bound states,
the Jost function, the scattering matrix,
the phase shift, and the spectral density associated with
(1.1) and (1.3); the exceptional and generic cases
that are related to $\lambda=0$ and $\lambda=4$
for the Schr\"odinger operator; Levinson's theorem;
and the Gel'fand-Levitan procedure associated with
(1.1) and (1.3).
In Section~3 we present the Darboux transformation formulas
when a bound state is added to the spectrum of the
Schr\"odinger operator. In Theorem~3.1 we prove that
the matrix inverses appearing in the relevant
Darboux transformation formulas in Section~3
 are well defined. In Section~4 we present the Darboux transformation formulas
when a bound state is removed from the spectrum of the
Schr\"odinger operator. In Theorem~4.1 we prove that
the matrix inverses appearing in the relevant
Darboux transformation formulas in Section~4
 are well defined. 
Finally, in Section~5 we present some illustrative examples for better understanding
of the results introduced
and also make a contrast between (1.1) and (1.2)
for certain results [1] related to compactly-supported
potentials.

The most relevant reference for our paper is [2], and in the current
paper we use the notation used in [2]. The results in [2]
were presented under the assumption that the potential is
compactly supported, i.e. $V_n=0$ for $n> b$ for some positive
integer $b.$ In Section~2 we present the corresponding results
when $V_n$ belongs to the Faddeev class and does not necessarily
have a compact support. Another relevant reference
for our paper is the classic work by Case and Kac [3].
Even though [3] is more related to
the Jacobi operator and not to the Schr\"odinger
operator, the treatment of the spectral density
in [3] is useful.
We remark that the Darboux transformation results related
to the Jacobi operators do not reduce to the Darboux transformation
results for the Schr\"odinger operator. Hence, in our
paper we use the Gel'fand-Levitan theory [3,4,10]
and an appropriate
formula for the spectral density for the
corresponding Schr\"odinger operator with bound states and derive the
Darboux transformation from first principles.

\vskip 5 pt
\noindent {\bf 2. PRELIMINARIES}
\vskip 3 pt

In this section, associated with (1.1) and (1.3)
we introduce various quantities such as
the Jost solution $f_n,$
the regular solution $\varphi_n,$ the Jost function
$f_0,$ the
scattering matrix $S,$ and the spectral measure
$d\rho.$ We also present the basic properties of
such quantities relevant to our analysis of Darboux transformations.

When the potential in (1.1) belongs to the
Faddeev class, the Schr\"odinger operator corresponding to (1.1) and
to the Dirichlet boundary condition (1.3) is a selfadjoint
operator acting on the class of square-summable functions. The spectrum of
the corresponding operator is well understood [2,3,6,7,11-13].
Let us use $\bR$ to denote the real
axis $(-\infty,+\infty).$
The continuous spectrum corresponds to
$\lambda\in[0,4],$ and the discrete spectrum
consists of at most a finite number of
discrete eigenvalues in $\bR\setminus[0,4],$ i.e.
$\lambda\in(-\infty,0)\cup (4,+\infty).$
For each $\lambda$-value in
the interval $(0,4),$ there are two linearly independent
solutions to (1.1).
There is only one linearly independent
solution satisfying both (1.1) and (1.3),
and such a solution is usually identified as a physical solution. Let us assume that
the discrete spectrum consists of $N$ eigenvalues
given by $\{\lambda_s\}_{s=1}^N,$ where $N=0$
corresponds to the absence of the discrete spectrum.
When $\lambda=\lambda_s,$ there is only one linearly independent
square-summable
solution satisfying (1.1) and (1.3).
For each of $\lambda=0$ and $\lambda=4,$ there exists one linearly independent solution
satisfying (1.1) and (1.3), and such a solution may be either bounded
in $n$ or it may grow as $O(n)$ as $n\to+\infty.$
For $\lambda=0,$ one says that the exceptional case occurs
if a solution satisfying (1.1) and (1.3) is bounded in $n$
and that the generic case occurs if a solution satisfying (1.1) and (1.3) is
not bounded in $n.$
Similarly, for $\lambda=4,$ the exceptional case occurs
if a solution satisfying (1.1) and (1.3) is bounded in $n$
and that the generic case occurs if a solution satisfying (1.1) and (1.3) is
not bounded in $n.$

In quantum mechanics, it is customary
to interpret the discrete spectrum
associated with (1.1) and (1.3) as the
bound states. Hence, the $\lambda_s$-values in the discrete spectrum can be called
the bound-state energies and the
corresponding square-summable
solutions can be called bound-state wavefunctions.
The solutions to (1.1) when $\lambda\in(0,4)$
can be referred to as scattering solutions.

Associated with (1.1), instead of $\lambda,$
it is convenient at times to use another
spectral parameter related to $\lambda,$ usually
denoted by $z,$ given by
$$z:=1-\displaystyle\frac{\lambda}{2}+\displaystyle\frac{1}{2}\,
\sqrt{\lambda(\lambda-4)},\tag 2.1$$
where the square root is used to denote the principal
branch of the complex
square-root function. Note that (2.1) yields
$$\lambda=2-z-z^{-1}.\tag 2.2$$
Let us use $\bold T$ for the unit
circle $|z|=1$ in the complex plane $\bold C,$ $\bold T^+$ for the upper
portion of $\bold T$ given by $z=e^{i\theta}$ with
$\theta\in(0,\pi),$ and $\overline{\bold T^+}$ for the closure
of $\bold T^+$ given by $z=e^{i\theta}$ with
$\theta\in[0,\pi].$
Under the transformation from $\lambda\in\bC$ to $z\in \bC,$
the real interval $\lambda\in(0,4)$ is mapped to
$z\in\bold T^+,$
the real half line $\lambda\in(-\infty,0)$ is mapped to the real interval
$z\in (0,1),$ the real interval $\lambda
\in(4,+\infty)$ is mapped to the real interval $z\in(-1,0),$ the point
$\lambda=0$ is mapped to $z=1,$ and
the point $\lambda=4$ is mapped to
$z=-1.$
Using (2.2) it is convenient to write (1.1) as
$$\psi_{n+1}+\psi_{n-1}=(z+z^{-1}+V_n)\,\psi_n,\qquad n\ge 1.\tag 2.3$$

Let us now consider certain particular solutions to
(1.1). A relevant solution to (1.1) or equivalently to (2.3)
is
the so-called regular solution $\varphi_n$ satisfying
the initial conditions
$$\varphi_0=0,\quad \varphi_1=1.\tag 2.4$$
 From (2.3) and (2.4) it follows that
$\varphi_n$ remains unchanged if we replace $z$ with $z^{-1}$ in $\varphi_n.$

The result presented in the following theorem is already known and its proof
is omitted. A proof in our own notation can be obtained as
in the proof of Theorem~2.6 of [2].

\noindent {\bf Theorem 2.1} {\it Assume that
the potential $V_n$ belongs to the Faddeev class. Then, for $n\ge 1$ the regular solution
$\varphi_n$ to (1.1) with the initial values (2.4) is a polynomial
in $\lambda$ of degree $n-1$ and is given by}
$$\varphi_n=\sum_{j=0}^{n-1} B_{nj} \lambda^j,\tag 2.5$$
{\it where, for each fixed positive integer $n,$
 the set of coefficients $\{B_{nj}\}_{j=0}^{n-1}$ are real valued and
uniquely determined by the ordered set
$\{V_1,V_2,\dots,V_{n-1}\}$ of potential values.
In particular, we have}
$$B_{n(n-1)}=(-1)^{n-1},\quad B_{n(n-2)}=(-1)^{n-2} \left[2(n-1)+\sum_{j=1}^{n-1} V_j
\right].$$

We remark that Theorem~2.1 holds even when
 the potential $V_n$ does not belong to the Faddeev class.
 If the potential values are allowed to be complex, then the coefficients
 $B_{nj}$ appearing in (2.5) are complex valued.

 From (2.5) it is clear that the $\lambda$-domain of $\varphi_n$ is the entire
 complex $\lambda$-plane. With the help of (2.2), we can conclude that
 the $z$-domain of $\varphi_n$ corresponds to the
 punctured complex $z$-plane with the point $z=0$ removed.

Another relevant solution to (1.1) or
equivalently to (2.3) is
the Jost solution $f_n$ satisfying
the asymptotic condition
$$f_n=z^n[1+o(1)],\qquad n\to+\infty.\tag 2.6$$
On the unit circle $z\in\bold T$ we have
$z^{-1}=z^\ast,$ where we use an asterisk to denote complex conjugation.
Let us use $f_n(z)$ to denote the value of $f_n$ when
$z\in\overline{\bold T^+}.$
 From (2.3) and (2.6) it follows that we have
$$f_n(z^{-1})=f_n(z^\ast)=f_n(z)^\ast,\qquad z\in\overline{\bold T^+},
\tag 2.7$$
and hence the domain of $f_n(z)$ can be extended
 from $z\in\overline{\bold T^+}$ to $z\in\bold T$
 by using (2.7). We will see in Theorem~2.2 that, when
 the potential $V_n$ belongs to the Faddeev class,
  the domain
 of $f_n(z)$ can be extended from $z\in\bold T$ to
 the unit disc $|z|\le 1.$

Let us define
$g_n$ as the quantity $f_n$ but by replacing $z$ by $z^{-1}$
there, i.e.
$$g_n(z):=f_n(z^{-1}),\qquad z\in\bold T.\tag 2.8$$
 From (2.8) it follows that
the domain of $g_n(z)$ is originally given as
$z\in\bold T$ and it can be extended to
$|z|\ge 1$ when the potential $V_n$ in (1.1)
belongs to the Faddeev class. With the help of (2.3) we see that
$g_n$ is also a solution to (1.1), and from (2.6) it follows that
$g_n$ satisfies the asymptotic condition
$$g_n=z^{-n}[1+o(1)],\qquad n\to+\infty.\tag 2.9$$

The quantity $f_0,$ which is obtained from the Jost solution $f_n$ with $n=0,$
is known as the Jost function.
Let us remark that the Jost solution $f_n$
is determined by the potential $V_n$ alone and
is unaffected by the choice of the Dirichlet
boundary condition (1.3).
On the other hand, the Dirichlet boundary condition (1.3)
is used when naming $f_0$ as the Jost
function. For a non-Dirichlet boundary condition
the Jost function is not defined as $f_0$
and it corresponds to an appropriate
linear combination of $f_0$ and
$f_1.$ In this paper we do not deal with
the Jost function in the non-Dirichlet case.

The Jost function $f_0(z)$ is used to define
the scattering matrix $S$ as
$$S(z):=\ds\frac{f_0(z)^\ast}{f_0(z)},\qquad z\in\bold T.
\tag 2.10$$
Even though $S(z)$ is scalar valued, it is customary to
refer to it as the scattering matrix.
With the help of (2.7) and (2.8) we see that we can write
(2.10) in various equivalent forms such as
$$S(z)=\ds\frac{g_0(z)}{f_0(z)}=\ds\frac{f_0(z^{-1})}{f_0(z)}
,\qquad z\in\bold T.
\tag 2.11$$

Let us write the Jost function in the polar form as
$$f_0(z)=|f_0(z)|\, e^{-i\,\phi(z)},\qquad z\in\bold T.
\tag 2.12$$
The real-valued quantity $\phi(z)$ appearing in (2.12) is usually called the
phase shift. Its domain consists of $z\in\bold T.$
Using (2.7) in (2.12) we see that the phase shift satisfies
$$\phi(z^{-1})=\phi(z^\ast)=-\phi(z),\qquad z\in\bold T.\tag 2.13$$
 From (2.10) we see that
the scattering matrix can be expressed in terms of the phase shift as
$$S(z)=e^{2i\,\phi(z)},\qquad z\in\bold T.
\tag 2.14$$

The relevant properties of the Jost solution $f_n$ and
the Jost function $f_0$
are summarized in the following theorem.

\noindent {\bf Theorem 2.2} {\it Assume that the potential
$V_n$ in (1.1) belongs to the Faddeev class.
Then:}

\item{(a)} {\it For each fixed $n=0,1,2,\dots,$ the Jost solution $f_n$ satisfying
(1.1) and (2.6) is analytic in $z$ in $|z|<1$ and continuous
in $z$ in $|z|\le 1.$ It has the representation}
$$f_n(z)=\sum_{m=n}^\infty K_{nm}\, z^m,\qquad |z|\le 1,\tag 2.15$$
{\it where each coefficients $K_{nm}$ is real valued
and uniquely determined by the potential values in the
ordered set $\{V_m\}_{m=n+1}^\infty.$
In particular, we have}
$$K_{nn}=1,\quad K_{n(n+1)}=\sum_{j=n+1}^\infty V_j,
\quad K_{n(n+2)}=\displaystyle\sum _{n+1\le j<l\le +\infty} V_j\, V_l.\tag
2.16$$

\item{(b)} {\it The Jost function $f_0$
is analytic in $z$ in $|z|<1$ and continuous
in $z$ in $|z|\le 1.$ It has the representation}
$$f_0(z)=\sum_{m=0}^\infty K_{0m}\, z^m,\qquad |z|\le 1,\tag 2.17$$
{\it where each coefficient $K_{0m}$ is
uniquely determined by the set $\{V_n\}_{n=1}^\infty$ of potential values.
In particular, we have}
$$K_{00}=1,\quad K_{01}=\sum_{j=1}^\infty V_j,
\quad K_{02}=\displaystyle\sum _{1\le j<l\le +\infty} V_j\, V_l.\tag
2.18$$

\item{(c)} {\it For each fixed $n=0,1,2,\dots,$ the solution $g_n$ satisfying
(1.1) and (2.9) is analytic in $z$ in $|z|>1$ and continuous
in $z$ in $|z|\ge 1.$ It has the representation}
$$g_n(z)=\sum_{m=n}^\infty K_{nm}\, z^{-m},\qquad |z|\ge 1.$$

\item{(d)} {\it The solutions $f_n$ and $g_n$ are linearly independent
when $z\in\bold T\setminus\{-1,1\}.$ In particular, the regular solution
$\varphi_n$ appearing in (2.4) can be expressed in terms of
$f_n$ and $g_n$ as}
$$\varphi_n=\displaystyle\frac{1}{z-z^{-1}}\,\left( g_0 f_n-f_0\, g_n\right).\tag 2.19$$

\noindent PROOF: It is enough to prove the analyticity
in $|z|<1$ and the continuity in $|z|\le 1$
for $f_n(z).$ The remaining results in (a)-(c)
can be obtained with the help of
Proposition~2.4 of [2]. Note that (2.19) is
the same as (2.42) of [2] and
the linear independence of $f_n$ and $g_n$ is established
by using (2.6) and (2.9).
Let us then prove the aforementioned
analyticity and continuity.
In fact, for the analyticity in $|z|<1,$ it is enough to
use the summability in (1.5) without the need for the first
moment of the potential. The first moment in (1.5) is needed
to prove the continuity at $z=\pm 1.$
We can prove the analyticity by modifying the proof of
Lemma~1 of [9] so that it is applicable to the discrete
Schr\"odinger equation.
We only provide the key steps and let the reader work out
the details.
Letting
$$m_n:= z^{-n} f_n,\tag 2.20$$
 from (2.6) we see that
$$m_n=1+o(1),\qquad n\to+\infty,$$
for each fixed $z\in\bold T.$
With the help of (2.3) and (2.20) we see that
$m_n$ satisfies the discrete equation given by
$$m_n=1+\ds\frac{1}{z-z^{-1}}\ds\sum_{j=n+1}^\infty \left(z^{2(j-n)}-1\right)\,V_j\,m_j.
\tag 2.21$$
Note that (2.21) is the discrete analog of
the second displayed formula on p. 130 of [9].
Next we solve (2.21) iteratively by letting
$$m_n(z)=\sum_{p=0}^\infty m_n^{(p)}(z),\qquad |z|<1,
\tag 2.22$$
where we have defined
$$m_n^{(0)}(z):=1,\qquad |z|<1,\tag 2.23$$
$$m_n^{(p)}(z):=\ds\frac{1}{z-z^{-1}}\ds\sum_{j=n+1}^\infty \left(z^{2(j-n)}-1\right)\,V_j\,m_j^{(p-1)}(z),
\qquad |z|<1, \quad p=1,2,3,\dots .\tag 2.24$$
Each iterate $m_n^{(p)}(z)$ is analytic in $|z|<1,$ and
the left-hand side of (2.22) is analytic in
$|z|<1$ if we can show that the series on the right-hand side of
(2.22) converges uniformly in every compact subset of
$|z|<1.$ When $|z|\le 1,$ we have
$$|z^{2(j-n)}-1|\le 2,\qquad j\ge n+1.\tag 2.25$$
Furthermore, from (1.5) we have
$$\ds\sum_{j=n+1}^\infty |V_j|\le \ds\sum_{j=1}^\infty |V_j|<+\infty.
\tag 2.26$$
The uniform convergence is established by using the
estimates in (2.25) and (2.26). Hence,
$m_n(z)$ is analytic in $|z|<1$ for each fixed nonnegative integer
$n.$
 From (2.20) it then follows that $f_n(z)$
 is analytic in $|z|<1$ for each fixed $n\ge 0.$
In order to prove the continuity of $m_n(z)$ in $|z|\le 1,$
 we need to show that
each iterate $m_n^{(p)}(z)$ is continuous in $|z|\le 1$ and
that the series in (2.22) converges absolutely
in $|z|\le 1.$ The factor $z-z^{-1}$ appearing in
the denominator of (2.24) becomes troublesome at
$z=\pm 1.$ As a remedy, we use the identity
$$\ds\frac{z^{2(j-n)}-1}{z-z^{-1}}=z\, \ds\frac{ z^{2j-2n}-1}{z^2-1}
=z\ds\sum_{k=0}^{j-n-1} z^{2k},
\qquad j\ge n+1.\tag 2.27$$
 From (2.27) it follows that for $|z|\le 1$ we have
$$\bigg| \ds\frac{z^{2(j-n)}-1}{z-z^{-1}}\bigg|\le j-n, \qquad j\ge n+1.
\tag 2.28$$
With the help of (1.5), (2.23), (2.24), and (2.28),
one establishes the uniform convergence in $|z|\le 1$
for the series on the right-hand side of
(2.22). Furthermore, with the help of
(2.24) and (2.27) we establish the continuity
of each iterate $m_n^{(p)}(z)$ in $|z|\le 1.$
Then, it follows that $m_n(z)$ appearing on the left-hand side
(2.22) is continuous in $|z|\le 1.$ Finally, from
(2.20) it follows that $f_n(z)$ is continuous
in $|z|\le 1$ for each fixed value of $n.$
\qed

Let us remark that, from (2.17) and (2.18) we see that the value of
the Jost function $f_0(z)$ at $z=0$ is given by
$$f_0(0)=1.\tag 2.29$$
 From the second equality of (2.16) it follows that
$$V_n=K_{(n-1)n}-K_{n(n+1)}, \qquad n=1,2,\dots.$$

The results in following theorem clarifies the generic
and exceptional cases encountered at the endpoints of
the continuous spectrum, i.e. at $\lambda=0$ and $\lambda=4.$

\noindent {\bf Theorem 2.3} {\it Assume that the potential $V_n$
appearing in (1.1)
belongs to the Faddeev class.
 Let $\lambda$ and
$z$ be the spectral parameters appearing in (1.1) and (2.1),
respectively, and let $\varphi_n$ and $f_n$ be the corresponding
regular solution and the Jost solution to (1.1)
appearing in (2.4) and (2.6), respectively.
Let $f_0$ be the corresponding Jost function.
Then:}

\item{(a)} {\it The Jost function $f_0(z)$ is nonzero when
$z\in\bold T\setminus\{-1,1\}.$}

\item{(b)} {\it At $\lambda=0,$ or equivalently at $z=1,$ the
regular solution $\varphi_n$ either grows linearly in $n$ as $n\to+\infty,$
which corresponds to the generic case, or it is bounded in $n,$ which
corresponds to the exceptional case. Hence, $\lambda=0$ never
corresponds to a bound state for (1.1) with the Dirichlet boundary
condition (1.3). In the generic case, $f_0\ne 0$ at $z=1.$
In the exceptional case, $f_0$ has a simple zero at $z=1.$}

\item{(c)} {\it At $\lambda=4,$ or equivalently at $z=-1,$ the
regular solution $\varphi_n$ generically grows linearly in $n$
as $n\to+\infty,$ and in the exceptional case the regular solution
$\varphi_n$ is bounded in $n.$ Hence,
$\lambda=4$ never corresponds to a bound state for
(1.1) with the Dirichlet boundary condition (1.3).
In the generic case we have $f_0\ne 0$ at $z=-1.$
In the exceptional case, $f_0$ has a simple zero at $z=-1.$}

\noindent PROOF: The proofs (b) and (c)
can be obtained as in the proof of
Theorem~2.5 of [2]. The proof of (a) can be given as follows.
Assume that $f_0(z)$ vanishes at some point $z=z_0,$ where
$z_0$ is located on the unit circle $\bold T$ and
$z_0\ne \pm 1.$ From (2.7) and (2.8) it follows that
$f_0(z_0)=0$ implies that
$g_0(z_0)=0.$ Using these values in (2.19) we would then get
$\varphi_n\equiv 0$ for any positive integer
$n$ when $z=z_0.$ On the other hand, by the second
equality in (2.4) we know that $\varphi_1$ must be equal to
$1$ when $z=z_0.$ This contradiction shows that
$f_0$ cannot vanish on the unit circle, except perhaps
at $z=\pm 1.$ \qed

The following theorem shows that the Jost function $f_0(z)$ cannot
vanish at any $z$-value inside the unit circle when
the imaginary part of that $z$-value is nonzero.

\noindent {\bf Theorem 2.4} {\it Assume that the potential $V_n$
appearing in (1.1)
belongs to the Faddeev class.
 Let $z$ be the spectral parameters appearing in (2.1),
$f_n(z)$ be the corresponding Jost solution
appearing in (2.15), and
$f_0(z)$ be the corresponding Jost function appearing in (2.17).
Then, $f_0(z)\ne 0$ for any $z$ satisfying $|z|<1$ with
the imaginary part
$\text{Im}[z]$ is nonzero. The zeros of $f_0(z)$ in the
interior of the unit circle can only occur when
$z\in(-1,0)\cup(0,1).$}

\noindent PROOF: From (2.17) we see that $f_0(0)=1,$ and hence it
is enough to prove that $f_0(z)\ne 0$ when
$|z|<1$ with $k_I\ne 0,$ where we use
the decomposition $z:=z_R+i\,z_I,$ with
$z_R$ and $z_I$ denoting the real and imaginary parts of $z,$ respectively.
For simplicity, let us use $f_n$ to denote $f_n(z).$ Since $f_n$ satisfies
(2.3) we have
$$f_{n+1}+f_{n-1}=(z+z^{-1}+V_n)\,f_n,\qquad n\ge 1.\tag 2.30$$
Taking the complex conjugate of both sides of
(2.30) and using the fact that $V_n$ is real, we obtain
$$f_{n+1}^\ast+f_{n-1}^\ast=\left[z^\ast+(z^\ast)^{-1}+V_n\right]\,f_n^\ast,\qquad n\ge 1.\tag 2.31$$
Let us multiply both sides of
(2.30) with $f_n^\ast$ and multiply both sides of
(2.31) with $f_n$ and subtract the resulting equations
side by side. This yields
$$f_n^\ast\,f_{n+1}+f_n^\ast\,f_{n-1}-
f_{n+1}^\ast\,f_n-f_{n-1}^\ast\,f_n
=\left[z-z^\ast+z^{-1}-(z^\ast)^{-1}\right]\,|f_n|^2,\qquad n\ge 1
.\tag 2.32$$
Note that
$$\text{Im}[z^{-1}]=\text{Im}\left[\ds\frac{1}{z_R+i\,z_I}\right]=\ds\frac{-z_I}{
z_R^2+z_I^2}.\tag 2.33$$
We have
$$z-z^\ast+z^{-1}-(z^\ast)^{-1}=2i\,\text{Im}[z]+2i\,\text{Im}[z^{-1}],\tag 2.34$$
and
using (2.33) in (2.34) we obtain
$$z-z^\ast+z^{-1}-(z^\ast)^{-1}=2i\,z_I-2i\,\ds\frac{z_I}{
z_R^2+z_I^2},$$
or equivalently
$$z-z^\ast+z^{-1}-(z^\ast)^{-1}=2i\,z_I\,\ds
\frac{z_R^2+z_I^2-1}{z_R^2+z_I^2}.\tag 2.35$$
Let us take the summation over $n$ on both sides of (2.32)
and use (2.35) in the resulting
summation, which
yields
$$\ds\sum_{n=1}^\infty \left[f_n^\ast\,f_{n+1}-f_{n-1}^\ast\,f_n\right]+
\ds\sum_{n=1}^\infty \left[f_n^\ast\,f_{n-1}-f_{n+1}^\ast\,f_n\right]=
2i\,z_I\,\ds
\frac{z_R^2+z_I^2-1}{z_R^2+z_I^2}\,\ds\sum_{n=1}^\infty |f_n|^2.\tag 2.36
$$
When $|z|<1,$ the two series on the left-hand side of (2.36) are both telescoping,
and using (2.6) in (2.36) we obtain
$$-f_0^\ast\,f_1+f_1^\ast\,f_0=
-2i\,z_I\,\ds
\frac{1-|z|^2}{|z|^2}\,\ds\sum_{n=1}^\infty |f_n|^2.\tag 2.37
$$
When $|z|<1$ with $z_I\ne 0,$ the right-hand side of
(2.37) cannot vanish unless
$f_n(z)=0$ for $n\ge 1.$ However, because of
(2.6) we cannot have $f_n(z)=0$ for all $n\ge 1$
at such a $z$-value.
Thus, the right-hand side of (2.37) cannot be zero
for any $z$-value satisfying $|z|<1$ with $z_I\ne 0.$ On the other hand,
if we had $f_0(z)=0$ for some $z$-value satisfying $|z|<1$ with $z_I\ne 0,$
then we would also have
$f_0(z)^\ast=0$ at the same $z$-value, and hence we would
have the left-hand side of (2.37) vanishing
at that $z$-value. This contradiction proves
that $f_0(z)\ne 0$ for any $z$-value satisfying $|z|<1$ with $z_I\ne 0.$
Since we have already seen that $f_0(0)\ne 0,$ we conclude that
the zeros of $f_0(z)$ in the interior of the unit circle
can only occur when $z\in(-1,0)\cup(0,1).$ \qed

In the next theorem, we summarize the facts relevant to the bound states of
(1.1) with the Dirichlet boundary condition (1.3). Recall that the
bound states correspond to the $\lambda$-values at which (1.1)
has square-summable solutions satisfying the boundary condition (1.3).

\noindent {\bf Theorem 2.5} {\it Assume that the potential $V_n$
appearing in (1.1)
belongs to the Faddeev class. Let $\lambda$ and
$z$ be the spectral parameters appearing in (1.1) and (2.1),
respectively, and let $f_n,$ $\varphi_n,$ and $f_0$ be the
corresponding Jost solution
appearing in (2.6), the regular solution appearing in (2.4),
and the Jost function appearing in (2.12), respectively. Then:}

\item{(a)} {\it A bound state can only occur when $\lambda\in (-\infty,0)$
or $\lambda\in (4,+\infty).$ Equivalently, a bound state can only occur
when $z\in(-1,0)$ or $z\in (0,1).$}

\item{(b)} {\it At a bound state, $\varphi_n$ and $f_n$ are both real valued for
every $n\ge 1.$ At a bound state, $\varphi_n$ and $f_n$ are linearly dependent
and each is square summable in $n.$}

\item{(c)} {\it At a bound state the Jost
function $f_0$ has a simple zero in $\lambda$ and
in $z.$ At a bound state the value of
the Jost solution at $n=1$ cannot vanish, i.e. $f_1\ne 0$
at a bound state.}

\item{(d)} {\it The number of bound states, denoted by $N,$ is finite. In particular,
we have $N=0$
when $V_n\equiv 0.$}

\noindent PROOF: The proofs for (a)-(c) can be obtained by slightly
modifying the proof
of Theorem~2.5 of [2] as follows. At a bound state,
(1.1) must have a square-summable solution satisfying the
Dirichlet boundary condition (1.3). Note that
(1.1) has two linearly independent solutions, and only one of
those two linearly independent solutions can satisfy (1.3). We know from the first equality
in (2.4) that the regular solution $\varphi_n$
appearing in (2.5) satisfies (1.3).
Thus, any bound-state
solution to (1.1) must be linearly dependent on $\varphi_n.$
Since the corresponding Schr\"odinger operator is selfadjoint,
the bound states can only occur when the spectral parameter
$\lambda$ is real. From (2.1) we know that
the $\lambda$-values in the interval
$\lambda\in(0,4)$ correspond to the $z$-values on
$\bold T^+,$ the upper portion of the
unit circle $\bold T.$ For such $z$-values, from (2.6) and (2.9) we
conclude that neither of the two linearly independent solutions
$f_n$ and $g_n$ can vanish as $n\to +\infty,$
where we recall that $g_n$ is the solution
appearing in (2.8). Furthermore, by (b) and (c)
of Theorem~2.3 we know that neither
$\lambda=0$ nor $\lambda=4$ can correspond to a bound state.
Thus, the bound states can only occur when
$\lambda\in (-\infty,0)$ or $\lambda\in(4,+\infty).$
Equivalently, with the help of (2.1) we conclude that
a bound state can only occur when $z\in(-1,0)$ or
$z\in(0,1).$ This completes the proof of
(a). Let us now prove (b). From Theorem~2.1 we know that the
coefficients $B_{nj}$ appearing in (2.5) are real valued,
and hence (2.5) implies that at any $\lambda$-value
in the interval
$\lambda\in(-\infty,0)$ or $\lambda\in(4,+\infty)$ the
corresponding
$\varphi_n$ is real valued for every $n\ge 1.$ Similarly, we know
 from Theorem~2.2(a) that the coefficients $K_{nm}$
 appearing in (2.15) are real valued, and hence
 (2.15) implies that $f_n$ for every $n\ge 1$ is
 real valued at any $z$-value occurring in
 $z\in(-1,0)\cup(0,1).$
 In the proof of (a) we have already indicated
 the linear dependence of
 $\varphi_n$ and $f_n$ and we have also indicated
that their square integrability follows from the definition of
a bound-state solution. Thus, the proof of (b) is complete.
Let us now turn to the proof of (c).
This follows by proceeding as in (2.67)-(2.69) of
[2] and hence by concluding that at a bound state the
Jost function $f_0$ must have a simple zero in
$\lambda$ and a simple zero in $z$ and that
$f_1$ cannot vanish at a bound state. This concludes the
proof of (c). Let us now prove (d).
We can see the finiteness
of the number of bound states as follows. From
Theorem~2.2 we know that $f_0(z)$ is analytic in $|z|<1$ and
continuous in $|z|\le 1.$ From (2.29) we know that
$f_0(0)=1.$ Furthermore, from (a) and (c) above
we know that
the bound states can only occur at the
zeros of $f_0(z)$ when $z\in(-1,0)\cup(0,1)$ and such zeros are simple.
Thus, the bound-state
zeros of $f_0(z)$ could only accumulate at $z=\pm 1.$
On the other hand, Theorem~2.3 indicates that $f_0(z)$
can at most have simple zeros at $z=\pm 1.$ Thus, $f_0(z)$
is analytic in $z\in(-1,1)$ with no accumulation points
in $z\in[-1,1].$ Consequently,
the number of bound-state zeros of $f_0(z)$ must be finite. \qed

For further elaborations on the finiteness of the
number of bound states, we refer the reader to [6,7] and the references
therein.

Let us assume that the bound states occur at $\lambda=\lambda_s$ for
$s=1,\dots,N.$ Let us also assume that the corresponding $z_s$-values
are obtained via using (2.1), and hence the bound states
occur at $z=z_s$ for $s=1,\dots,N.$ From (2.2) we see that
$$\lambda_s=2-z_s-z_s^{-1},\qquad s=1,\dots,N.\tag 2.38$$
 From Theorem~2.5(b)
we know that $\varphi_n(\lambda_s)$ is real valued and the
quantity $C_s$ defined as
$$C_s:=\ds\frac{1}{\sqrt{\ds\sum_{n=1}^\infty \varphi_n(\lambda_s)^2}},
\qquad s=1,\dots,N,\tag 2.39$$
is a finite nonzero number. It is appropriate to refer to
the positive number $C_s$ as the Gel'fand-Levitan norming constant
at $\lambda=\lambda_s.$
Thus, the quantity $C_s \varphi_n(\lambda_s)$ is a normalized
bound-state solution to (1.1) at the bound state $\lambda=\lambda_s.$
Similarly, from Theorem~2.5(b)
we know that $f_n(z_s)$ is real valued and the
quantity $c_s$ defined as
$$c_s:=\ds\frac{1}{\sqrt{\ds\sum_{n=1}^\infty f_n(z_s)^2}},
\qquad s=1,\dots,N,\tag 2.40$$
is a finite nonzero number. It is appropriate to refer to
the positive number $c_s$ as a Marchenko norming constant
at $z=z_s.$
Thus, the quantity $c_s f_n(z_s)$ is a normalized
bound-state solution to (1.1) at the bound state $z=z_s.$
We then get
$$C_s^2 \left[\varphi_n(\lambda_s)\right]^2=
c_s^2\left[ f_n(z_s)\right]^2,\qquad s=1,\dots,N.\tag 2.41$$
 Using the second equality of (2.4) in (2.41) we see that
 the Gel'fand-Levitan norming constant
 $C_s$ and the Marchenko norming constant $c_s$ are related to each other as
$$C_s^2=c_s^2  \left[ f_1(z_s)\right]^2,\qquad s=1,\dots,N.\tag 2.42$$

Let us use a circle above a quantity to emphasize that
it corresponds to the trivial potential
$V_n\equiv 0$ in (1.1). Hence, $\overset{\circ}\to{\varphi}_n$
denotes the regular solution,
$\overset{\circ}\to f_n$ is the Jost solution,
$\overset{\circ}\to g_n$ is related to
$\overset{\circ}\to f_n$ as in
(2.8), $\overset{\circ}\to f_0$ is
the Jost solution, and
$\overset{\circ}\to S$ is the scattering matrix.
We have [2]
$$\overset{\circ}\to f_n=z^n,\quad \overset{\circ}\to g_n=z^{-n},
\quad \overset{\circ}\to{\varphi}_n=\ds\frac{z^n-z^{-n}}{z-z^{-1}},
\qquad
n\ge 1,$$
$$\overset{\circ}\to f_0(z)\equiv 1,\quad
\overset{\circ}\to g_0(z)\equiv 1,\quad \overset{\circ}\to S(z)\equiv 1.$$

Let us use $d\rho$ to denote the spectral
density corresponding to the Schr\"odinger equation
(1.1) with the Dirichlet boundary condition (1.3).
The spectral density
is normalized, i.e. its integral over the real-$\lambda$ axis
is equal to one.
Let us use $d\overset{\circ}\to \rho$ to denote the spectral
density when the potential is zero.
 From (4.1) of [2] we have
$$d\overset{\circ}\to\rho=\cases 0,\qquad \lambda <0,\\
\noalign{\medskip}
\displaystyle\frac{1}{2\pi}\,\sqrt{\lambda(4-\lambda)}\,d\lambda,\qquad 0\le \lambda\le 4,\\
\noalign{\medskip} 0,\qquad \lambda> 4.\endcases\tag 2.43$$
 From (2.43) we see that, when the potential is zero, the discrete
 part of the spectral measure, i.e. the part
 corresponding to
 $\bR\setminus[0,4]$ is zero. Thus, the continuous
 part of the spectral density in (2.43) has its integral
 over $\lambda\in[0,4]$ equal to one. Using (2.2) in (2.43), we can express [2]
 the continuous part of $d\overset{\circ}\to\rho$ in terms of $z$ as
 $$d\overset{\circ}\to\rho=-\displaystyle\frac{1}{2\pi i}\,(z-z^{-1})^2\,\displaystyle\frac{dz}{z},
\qquad z\in \overline{\bold T^+},$$
where we recall that
$\overline{\bold T^+}$ denotes the closure of the upper portion of
the unit circle $\bold T.$

 In the absence of bound states, the spectral density $d\rho$ is given
 by
 $$d\rho=\cases \ds\frac{d\overset{\circ}\to\rho}{|f_0(z)|^2},\qquad \lambda\in[0,4],\\
 \stretch
 0,\qquad\lambda\in\bR\setminus[0,4],\endcases\tag 2.44$$
 where we recall that $\lambda\in[0,4]$ corresponds to
 $z\in \overline{\bold T^+}.$
Thus, the discrete part of the spectral density
$d\rho$ is zero and the continuous part of the
spectral density is obtained by
dividing $d\overset{\circ}\to\rho$ by the absolute
square of the Jost function
$f_0(z).$ When there are
$N$ bound states at $\lambda=\lambda_s$
with the corresponding Gel'fand-Levitan
norming constants $C_s$ appearing in (2.39),
one can evaluate the spectral density $d\rho$ as
$$d\rho=\cases \ds\frac{1-\ds\sum_{s=1}^N C_s^2}{
\ds\prod_{k=1}^N z_k^2}\,\ds\frac{ d\overset{\circ}\to\rho}{|f_0(z)|^2},
\qquad \lambda\in[0,4],\\
\stretch
\ds\sum_{s=1}^N C_s^2\,\delta(\lambda-\lambda_s)\, d\lambda,\qquad
\lambda\in\bR\setminus[0,4],\endcases\tag 2.45$$
where $f_0(z)$ is the corresponding Jost function
and each $z_s$ corresponds to $\lambda_s$ via (2.38).
We remark that $\lambda\in[0,4]$ in (2.45) corresponds to
 $z\in \overline{\bold T^+}.$
Note that, in the absence of
bound states, i.e. when $N=0,$ the spectral density
given in (2.45) reduces to the expression given in (2.44).
In the evaluation of (2.45) we have used the facts that
$$\ds\int_{\lambda\in\bR} d\rho=1,\quad
\int_{\lambda\in\bR\setminus[0,4]} d\rho=\ds\sum_{s=1}^N C_s^2,\quad
 \int_{\lambda\in[0,4]} d\rho=1-\ds\sum_{s=1}^N C_s^2.\tag 2.46$$
With the help of (2.46) we see that the first line of (2.45) yields
$$\ds\ds\int_{\lambda\in[0,4]} \ds\frac{ d\overset{\circ}\to\rho}{|f_0(z)|^2}=
\prod_{k=1}^N z_k^2.$$

In order to understand the Darboux transformation,
 we need to establish the Gel'fand-Levitan formalism related to (1.1) and (1.3).
Let $V_n$ and $\tilde V_n$ be the unperturbed and perturbed
potentials, respectively. Let $\varphi_n$ and $\tilde\varphi_n$ be the respective
corresponding regular solutions, and let $d\rho$ and $d\tilde\rho$ be the
respective corresponding
spectral densities. From Theorem~2.1 it follows that
$$\tilde \varphi_n=\cases \varphi_n,\qquad n=1,\\
\stretch
\varphi_n+\ds\sum_{m=1}^{n-1} A_{nm}\,\varphi_m,\qquad
n=2,3,\dots,\endcases\tag 2.47$$
where $A_{nm}$ are some real coefficients to be determined.
Let us define the real-valued scalars $G_{nm}$ as
$$G_{nm}:=\int_{\lambda\in\bR} \varphi_n\, [d\tilde\rho-d\rho]\,\varphi_m.\tag 2.48$$
We already have [2,3] the orthonormality
$$\int_{\lambda\in\bR} \varphi_n\, d\rho\,\varphi_m=\delta_{nm},\tag 2.49$$
with $\delta_{nm}$ denoting the Kronecker delta.
Proceeding as in (4.13)-(4.17) of [2] we obtain the
Gel'fand-Levitan system
$$A_{nm}+G_{nm}+\ds\sum_{j=1}^{n-1} A_{nj} G_{jm}=0,\qquad 1\le m<n.
\tag 2.50$$
Analogous to (2.84) of [2], we get
$$\tilde V_n-V_n=A_{(n+1)n}-A_{n(n-1)},\qquad n=1,2,3,\dots,
\tag 2.51$$
with the understanding that $A_{10}=0.$

For each integer $n\ge 2,$ let $\bold G_{n-1}$ be the $(n-1)\times (n-1)$ matrix whose $(k,l)$-entry
is equal to $G_{kl}$ evaluated as in (2.48), i.e.
$$\bold G_{n-1}:=\bm G_{11}&G_{12}&\cdots & G_{1(n-2)}&G_{1(n-1)}\\
G_{21}&G_{22}&\cdots & G_{2(n-2)}&G_{2(n-1)}\\
\vdots& \vdots& \ddots& \vdots&\vdots\\
G_{(n-2)1}&G_{(n-2)2}&\cdots & G_{(n-2)(n-2)}&G_{(n-2)(n-1)}\\
G_{(n-1)1}&G_{(n-1)2}&\cdots & G_{(n-1)(n-2)}&G_{(n-1)(n-1)}\endbm.\tag 2.52$$
 From (2.48) and (2.52) we see that
$\bold G_{n-1}$ is a real symmetric matrix.
For each integer $n\ge 2,$ we can write
the Gel'fand-Levitan system (2.50) in the matrix form as
$$(I_{n-1}+\bold G_{n-1})
\bm A_{n1}\\
A_{n2}\\
\vdots\\
A_{n(n-2)}\\
A_{n(n-1)}\endbm=-\bm G_{n-1}\\
G_{n2}\\
\vdots\\
G_{n(n-2)}\\
G_{n(n-1)}\endbm,\tag 2.53$$
where $I_{n-1}$ is the $(n-1)\times (n-1)$ identity matrix.
Let $\bold g_{n-1}$ be the column vector with $(n-1)$ components
appearing on the right-hand side of (2.53), i.e.
$$\bold g_{n-1}:=\bm G_{n1}&G_{n2}&\cdots&G_{n(n-2)}&G_{n(n-1)}\endbm^\dagger.
\tag 2.54$$
Using (2.54) in (2.53) we obtain
$$\bm A_{n1}\\
A_{n2}\\
\vdots\\
A_{n(n-2)}\\
A_{n(n-1)}\endbm=-(I_{n-1}+\bold G_{n-1})^{-1}\,\bold g_{n-1}.\tag 2.55$$
Thus, $A_{nm}$
can be explicitly expressed in terms
of the coefficients of $\bold G_{n-1}$ as
$$A_{nm}=-\hat 1_m^\dagger\,
(I_{n-1}+\bold G_{n-1})^{-1}\,\bold g_{n-1},\qquad 1\le m<n,\tag 2.56$$
where $\hat 1_m$ is the column vector with $(n-1)$ components with all
the entries being zero except for the $m$th entry being one.
Note that the right-hand side of (2.56) contains a binomial form for
a matrix inverse.
Using (15) on p. 15 of [5], the binomial
form in (2.56) can be expressed as a ratio of
two determinants, yielding
$$A_{nm}=\ds\frac{\det\bm 0& \hat 1_m^\dagger\\
\stretch
\bold g_{n-1}&
(I_{n-1}+\bold G_{n-1})\endbm}{\det[I_{n-1}+\bold G_{n-1}]},\qquad 1\le m<n,\tag 2.57$$
where the matrix in the numerator is a block matrix of size $n\times n.$
Using (2.57) in (2.47) and (2.51) we obtain
$\tilde \varphi_n$ and $\tilde V_n$ in terms of the
unperturbed quantities.

Let us refer to the data set $\{\lambda_s,C_s\}_{s=1}^N,$ which consists of
all the bound-state energies and the corresponding
Gel'fand-Levitan norming
constants given in (2.39), as the bound-state data set.
In general, the scattering matrix $S(z)$ defined in (2.10)
and the bound-state data set are independent. This is because
the domain of $S(z)$ consists of the unit circle
$z\in\bold T$ and the bound-state energies
correspond to the $z_s$-values inside the unit circle.
Let us consider the case where the nontrivial
potential $V_n$ is compactly supported, i.e.
when $V_n=0$ for $n>b$ and $V_b\ne 0$ for some positive integer
$b.$ Thus, we use $b$ to signify the compact support of
$V_n$ given by $\{1,2,\dots,b\}.$ For such potentials, it is known [2] that $S(z)$ has a meromorphic extension from
$z\in\bold T$ to the region $|z|<1$ and
the $z_s$-values correspond to the poles of $S(z)$ in $|z|<1.$
Furthermore, for such potentials the
corresponding $C_s$-values can be determined [2]
in terms of certain residues evaluated at $z_s$-values.
In general, without a compact support,
the values of $z_s$ and $C_s$ cannot be determined
 from the scattering matrix $S(z).$ On the other hand,
 even without a compact support,
when the potential $V_n$ belongs to the Faddeev class,
the scattering matrix corresponding (1.1) and (1.3)
contains some information related to the
number of bound states $N.$ This result is known
as Levinson's theorem, and mathematically it can be viewed
as an argument principle related
to the integral of the logarithmic derivative
of the scattering matrix along the unit circle $\bold T$ in the complex
$z$-plane.

In the next theorem, we present Levinson's theorem
associated with (1.1) and (1.3). For this purpose it is appropriate
to introduce the constants $\mu_+$ and $\mu_-$ as
$$\mu_+:=\cases 1,\qquad f_0(1)=0,\\
\stretch
0,\qquad f_0(1)\ne 0,\endcases\tag 2.58$$
$$\mu_-:=\cases 1,\qquad f_0(-1)=0,\\
\stretch
0,\qquad f_0(-1)\ne 0.\endcases\tag 2.59$$
Let us elaborate on (2.58) and (2.59).
 From Theorem~2.3(b), we know that
 $\mu_+=1$ if we have the exceptional case at $z=1$ and
 we have $\mu_+=0$ if we have the generic
 case at $z=1.$ Similarly,
 from (2.59) and Theorem~2.3(c) we conclude
 that $\mu_-=1$ if we have the exceptional case at $z=-1$ and
 we have $\mu_-=0$ if we have the generic
 case at $z=-1.$

 Let
$\Delta_{\bold T}$ acting on a function of $z$ denote the change
in that function
when the $z$-value moves along
the unit circle
$\bold T$ once in the counterclockwise
direction in the sense of the Cauchy principal value. By the sense of the Cauchy principal
value, we mean that in the evaluation of
the change by using an integral along $\bold T,$
we interpret the corresponding integral as a Cauchy principal value. In the
theorem given below, that amounts to integrating along the
unit circle $z=e^{i\theta}$ for $\theta\in(0^+,\pi-0^+)\cup(\pi+0^+,2\pi-0^+)$
because the only singularities for the integrand may occur at
$z=1$ or $z=-1.$

\noindent {\bf Theorem 2.6} {\it Assume that the potential $V_n$
appearing in (1.1)
belongs to the Faddeev class. Let $f_0(z)$ appearing
in (2.12), $S(z)$ appearing in (2.10), $\phi(z)$
appearing in (2.12), and $N$ appearing in (2.39)
be the respective Jost function, the scattering matrix, the phase shift,
and the number of bound states
corresponding to (1.1) and (1.3). Let
$\Delta_{\bold T}$ signify the change
when the $z$-value moves along
the unit circle
$\bold T$ once in the counterclockwise
direction in the sense of the Cauchy principal value. We then have the following:}

\item{(a)} {\it The change in the phase shift $\phi(z)$ when
$z$ moves along $\bold T$ in the counterclockwise direction once
is given by}
$$\Delta_{\bold T} [\phi(z)]=-\pi\left[2N+\mu_++\mu_-\right],
\tag 2.60$$
{\it where $\mu_+$ and $\mu_-$ are the constants
defined in (2.58) and (2.59), respectively.}

\item{(b)} {\it The change in the phase shift $\phi(z)$ when
$z$ moves along $\bold T^+$ from $z=1$ to $z=-1$
is given by}
$$\Delta_{\bold T^+} [\phi(z)]=-\pi\left[N+\ds\frac{\mu_+}{2}+\ds\frac{\mu_-}{2}\right].
\tag 2.61$$

\item{(c)} {\it The change in the argument of $S(z)$ when
$z$ moves along $\bold T^+$ from $z=1$ to $z=-1$
is given by}
$$\Delta_{\bold T^+} [\text{arg}[S(z)]]=-\pi\left[2N+\mu_++\mu_-\right].
\tag 2.62$$

\item{(d)} {\it The change in the argument of $f_0(z)$ when
$z$ moves along $\bold T^+$ from $z=1$ to $z=-1$
is given by}
$$\Delta_{\bold T^+} [\text{arg}[f_0(z)]]=\pi\left[N+\ds\frac{\mu_+}{2}+\ds\frac{\mu_-}{2}\right].
\tag 2.63$$

\noindent PROOF: From Theorem~2.2(b) we know that
$f_0$ is analytic in $|z|<1$ and continuous in $|z|\le 1.$
Thus, $f_0$ has no singularities in $|z|\le 1.$ On the other hand,
 from Theorem~2.4 and Theorem~2.5(c) we know that
 the only zeros of $f_0$ in $|z|<1$ occur
 at the bound states, those zeros are simple and
 can only occur when
 $z\in(-1,0)$ or $z\in(0,1),$ the number of
 such zeros is finite, and we use $N$ to denote the nonnegative integer
specifying the number of bound states. From Theorem~2.3 we know that
 the only zeros of $f_0$ on
 $z\in\bold T$ may occur at $z=\pm 1,$ such zeros are simple, and the number of
 such zeros is equal to $\mu_++\mu_-.$ Applying the
 argument principle to $f_0(z)$ along the unit circle, we see that
the change in the argument of $f_0(z)$ as $z$ moves
along the unit circle once in the counterclockwise direction
is given by
$$\Delta_{\bold T} [\text{arg}[f_0(z)]=2\pi\left[N+\ds\frac{\mu_+}{2}+\ds\frac{\mu_-}{2}\right],
\tag 2.64$$
where we have used the fact that the contribution from a zero of $f_0(z)$
on $z\in\bold T$ is half of the contribution from a zero
in $|z|<1.$ Using (2.12) and (2.64) we obtain
(2.60). Using (2.13) in (2.60) we obtain (2.61). Using
(2.14) in (2.61) we get (2.62). Using (2.13) in (2.64) we have
(2.63). \qed

\vskip 8 pt
\noindent {\bf 3. DARBOUX TRANSFORMATION IN ADDING A BOUND STATE}
\vskip 3 pt

In this section we determine the effect of adding a bound state
to the discrete spectrum of the Schr\"odinger operator
corresponding to (1.1) and (1.3). For clarity, we use the notation
$V_n(N)$ for $V_n$ to indicate that the Schr\"odinger operator
 contains exactly $N$ bound states occurring
at $\lambda=\lambda_s$ for $s=1,\dots,N.$ Hence, we order
the bound states by assuming that we start with
the potential $V_n(0)$ containing no bound states. Then, we add one bound state
at $\lambda=\lambda_1$ with some Gel'fand-Levitan norming constant and
obtain the potential $V_n(1).$ Next, we add one bound state
at $\lambda=\lambda_2$ with some
Gel'fand-Levitan norming constant and
obtain the potential $V_n(2).$ Continuing in this manner
we recursively add all the bound states with
$\lambda=\lambda_s$ for $s=1,\dots,N$ and obtain the potential $V_n(N).$
Note that (2.38) establishes a one-to-one correspondence between $\lambda_s$ and
$z_s,$ and hence we can equivalently say that the bound states of
the potential $V_n(N)$ occur at $z=z_s$
for $s=1,\dots,N.$ We remark that the ordering of $\lambda_s$ is completely
arbitrary and that ordering does not have to have $\lambda_s$
in an ascending or descending order.

To the ``unperturbed" potential $V_n(N)$ let us add one bound state
at $\lambda=\lambda_{N+1}$ with the Gel'fand-Levitan norming constant
$C_{N+1}.$ We then get the ``perturbed" potential $V_n(N+1).$
Equivalently stated, we add one bound states at $z=z_{N+1},$
where $z_{N+1}$ and $\lambda_{N+1}$ are related to each other via (2.38)
and $z_{N+1}\in (-1,0)\cup(0,1).$
The Jost function for the unperturbed problem is
denoted by $f_0(z;N)$ and the Jost function for the
perturbed problem is denoted by $f_0(z;N+1).$
In the analog of adding a bound state for the Schr\"odinger
equation (1.2), we can uniquely express the perturbed
Jost function in terms of the unperturbed Jost function
by requiring that the absolute value of the Jost function
in the continuous spectrum remains unchanged [4]. However,
this is no longer the case for the discrete
Schr\"odinger equation. Let us elaborate on this matter.
We would like $f_0(z;N+1)$ to be obtained from $f_0(z;N)$ via
$$f_0(z;N+1)=\left(1-\ds\frac{z}{z_{N+1}}\right)\,Q(z)\,f_0(z;N),\qquad |z|\le 1,
\tag 3.1$$
where $Q(z)$ is analytic in
$|z|<1,$ continuous in $|z|\le 1,$ and satisfies $Q(0)=1.$ The constraints
on $Q(z)$ are determined by the fact that
both $f_0(z;N+1)$ and $f_0(z;N)$ must be analytic
in $|z|<1,$ continuous in $|z|\le 1,$ and take the value of
$1$ at $z=0,$ as required by Theorem~2.2(b).
Furthermore, $f_0(z;N+1)$ must have a simple zero at
$z=z_{N+1}$ and $f_0(z;N)$ must be nonzero when $z=z_{N+1}.$
The further requirement
$$|f_0(z;N+1)|=|f_0(z;N)|,\qquad z\in\bold T,\tag 3.2$$
combined with the maximum modulus principle would yield
$$\left(1-\ds\frac{z}{z_{N+1}}\right)\,Q(z)\equiv 1,\qquad |z|\le 1.
\tag 3.3$$
The result in (3.3) would follow from the fact that an analytic
function in a bounded domain must take its maximum modulus value
somewhere on the boundary, and it can be obtained as follows.
The left-hand side of
(3.3) is already equal to one at the interior point $z=0$
and hence (3.3) must hold for all $z$-values on the unit
disk $|z|\le 1.$ On the other hand, (3.3) is not acceptable because it
requires $Q(z)$ to have a pole at $z=z_{N+1},$ contradicting
the requirement that $Q(z)$ is analytic in $|z|<1.$
Thus, in adding a bound state, we must use (3.1) without
requiring (3.2).

In establishing a Darboux transformation,
the choice of $Q(z)$ appearing in (3.1) is not unique.
We find it convenient to choose a particular
$Q(z)$ as
$$Q(z)=\ds\frac{1}{1-z_{N+1} z},\qquad |z|\le 1.
\tag 3.4$$
One could argue that the simplest choice
$Q(z)\equiv 1$ would be a better choice than the one given in
(3.4). It turns out that the choice
in (3.4) has a few important advantages
over other choices. For example, with the choice
of $Q(z)$ given in (3.4) we obtain
$$|f_0(z;N+1)|^2=\ds\frac{1}{z_{N+1}^2}\,|f_0(z;N)|^2,\qquad z\in\bold T,\tag 3.5$$
which greatly simplifies evaluations involving the spectral
density given in (2.45). On the other hand, the choice
$Q(z)\equiv 1$ yields
$$|f_0(z;N+1)|^2=\bigg|1-\ds\frac{z}{z_{N+1}}\bigg|^2\,|f_0(z;N)|^2,
\qquad z\in\bold T,$$
which hinders evaluations involving the spectral density.
Another advantage of the choice
of $Q(z)$ given in (3.4) is that the pole of $Q(z)$
at $z=1/z_{N+1}$ can be considered as a real-valued resonance
for the discrete Schr\"odinger equation (1.1), where we
recall that $z_{N+1}\in (-1,0)\cup (0,1).$
Consider the special case of a compactly-supported potential,
where
$z=z_{N+1}$ is a real-valued resonance
for $V_n(N),$ i.e.
$f_0(z;N)$ has a simple zero at $z=1/z_{N+1}.$
We may then be able to convert that resonance
to a bound state by adding a bound state to
$V_n(N)$ at $z=z_{N+1}$ in such a way that
$V_n(N+1)$ contains a bound state. We refer the reader to
[1], where the analogous problem for (1.2) of converting a resonance into
a bound state without affecting the compact support property of the potentials.
For the discrete Schr\"odinger operator associated with
(1.1) and (1.3), in some of the examples in Section~5
we illustrate converting a resonance into a bound state and determine how the
compact-support property is impacted.

In our paper we exclusively use the choice in (3.4) in adding a bound state.
Hence, as seen from (3.1) and (3.4), the Darboux transformation formula
for the Jost function in adding one bound state at $z=z_{N+1}$
with $z_{N+1}\in(-1,0)\cup (0,1)$ yields
$$f_0(z;N+1)=\ds\frac{1-\ds\frac{z}{z_{N+1}}}{1-z_{N+1}\,z}\,f_0(z;N),\qquad |z|\le 1.
\tag 3.6$$
Let $S(z;N)$ and $S(z;N+1)$ denote the scattering matrices for the
unperturbed and perturbed problems, respectively. From (2.11) we get
$$S(z;N)=\ds\frac{f_0(z^{-1};N)}{f_0(z;N)},\quad
S(z;N+1)=\ds\frac{f_0(z^{-1};N+1)}{f_0(z;N+1)},\qquad z\in\bold T.
\tag 3.7$$
Using (3.6) in (3.7), after some simplification, we obtain
the Darboux transformation for the scattering matrix as
$$S(z;N+1)=\left(\ds\frac{1-z_{N+1}\,z}{z-z_{N+1}}\right)^2\,S(z;N),\qquad z\in\bold T.
\tag 3.8$$
One can directly verify that
$$\bigg|\ds\frac{1-z_{N+1}\,z}{z-z_{N+1}}\bigg|^2=1,\qquad z\in\bold T,$$
and hence, with the help of (2.14), we see that
the Darboux transformation for the phase shift is given by
$$\phi(z;N+1)=\phi(z;N)-\ds\frac{i}{2}\,\log\left(\ds\frac{1-z_{N+1}\,z}{z-z_{N+1}}\right)^2,
\qquad z\in\bold T.
\tag 3.9$$

Next, let us determine the Darboux transformation for the spectral density.
Let $d\rho(\lambda;N)$ and $d\rho(\lambda;N+1)$ denote the unperturbed and
perturbed spectral densities, respectively.
 From (2.45) we see that
$$d\rho(\lambda;N)=\cases \ds\frac{1-\ds\sum_{s=1}^N C_s^2}{
\ds\prod_{k=1}^N z_k^2}\,\ds\frac{ d\overset{\circ}\to\rho}{|f_0(z;N)|^2},
\qquad \lambda\in[0,4],\\
\stretch
\ds\sum_{s=1}^N C_s^2\,\delta(\lambda-\lambda_s)\,d\lambda,\qquad
\lambda\in\bR\setminus[0,4],\endcases\tag 3.10$$
 $$d\rho(\lambda;N+1)=\cases \ds\frac{1-\ds\sum_{s=1}^{N+1} C_s^2}{
\ds\prod_{k=1}^{N+1} z_k^2}\,\ds\frac{ d\overset{\circ}\to\rho}{|f_0(z;N+1)|^2},
\qquad \lambda\in[0,4],\\
\stretch
\ds\sum_{s=1}^{N+1} C_s^2\,\delta(\lambda-\lambda_s)\,d\lambda,\qquad
\lambda\in\bR\setminus[0,4],\endcases\tag 3.11$$
where we recall that $\lambda\in[0,4]$ corresponds to
 $z\in \overline{\bold T^+}.$
Using (3.5) in (3.11) we see that
 $$d\rho(\lambda;N+1)=\cases \ds\frac{1-\ds\sum_{s=1}^{N+1} C_s^2}{
\ds\prod_{k=1}^N z_k^2}\,\ds\frac{ d\overset{\circ}\to\rho}{|f_0(z;N)|^2},
\qquad \lambda\in[0,4],\\
\stretch
\ds\sum_{s=1}^{N+1} C_s^2\,\delta(\lambda-\lambda_s)\,d\lambda,\qquad
\lambda\in\bR\setminus[0,4],\endcases\tag 3.12$$
 and hence
 from (3.10) and (3.12) we get the Darboux transformation for
 the spectral density as
$$ d\rho(\lambda;N+1)-d\rho(\lambda;N)=
\cases -\ds\frac{C_{N+1}^2}{1-\ds\sum_{s=1}^N C_s^2}
\,d\rho(\lambda;N),
\qquad \lambda\in[0,4],\\
\stretch
C_{N+1}^2\,\delta(\lambda-\lambda_{N+1})\,d\lambda,\qquad
\lambda\in\bR\setminus[0,4].\endcases\tag 3.13$$

Our next goal is to determine the Darboux transformation for the regular solution.
In other words, we would like to determine the relationship between
$\varphi_n(\lambda;N)$ and $\varphi_n(\lambda;N+1),$ where the former is
the regular solution for the unperturbed problem and the latter
is the regular solution for the perturbed problem.

Let us now use the Gel'fand-Levitan procedure
in the special case with $V_n(N+1)$ denoting
$\tilde V_n$ and $V_n(N)$ denoting $V_n.$ In that
special case $d\rho$ and $d\tilde \rho$
appearing in (2.47) correspond to
$d\rho(\lambda;N)$ and $d\rho(\lambda;N+1),$ respectively, appearing
on the left-hand side of (3.13). The unperturbed and perturbed
regular solutions
$\varphi_n$ and $\tilde\varphi_n$ appearing
in (2.47) correspond to
$\varphi_n(\lambda;N)$ and $\varphi_n(\lambda;N+1),$
respectively.
 From the second line of (3.10) we obtain
$$\int_{\lambda\in\bR\setminus[0,4]} \varphi_n(\lambda;N)\,d\rho(\lambda;N)\, \varphi_m(\lambda;N)=\sum_{s=1}^N C_s^2\,
\varphi_n(\lambda_s;N)\,\varphi_m(\lambda_s;N).\tag 3.14$$
With the help of (2.49) and (3.14) we get
$$\int_{\lambda\in[0,4]} \varphi_n(\lambda;N)\,d\rho(\lambda;N)\, \varphi_m(\lambda;N)=
\delta_{nm}-\sum_{s=1}^N C_s^2\,
\varphi_n(\lambda_s;N)\,\varphi_m(\lambda_s;N),\tag 3.15$$
where we recall that $\delta_{nm}$ denotes the Kronecker delta.
Using (3.13) in (2.48) we obtain
$$\aligned
G_{nm}=&-\ds\frac{C_{N+1}^2}{1-\ds\sum_{k=1}^N C_k^2}
\,\int_{\lambda\in [0,4]} \varphi_n(\lambda;N)\,d\rho(\lambda;N)\, \varphi_m(\lambda;N)\\
\stretch
&+C_{N+1}^2 \,\varphi_n(\lambda_{N+1};N)\,\varphi_m(\lambda_{N+1};N).
\endaligned\tag 3.16$$
The integral in (3.16) is equal to the right-hand side of (3.15).
Thus, from (3.15) and (3.16) we obtain
$$\aligned
G_{nm}=&-\ds\frac{C_{N+1}^2}{1-\ds\sum_{k=1}^N C_k^2}\,
\delta_{nm}+\ds\frac{C_{N+1}^2}{1-\ds\sum_{k=1}^N C_k^2}
\,\ds\sum_{s=1}^N C_s^2\,
\varphi_n(\lambda_s;N)\,\varphi_m(\lambda_s;N)\\
&+C_{N+1}^2 \,\varphi_n(\lambda_{N+1};N)\,\varphi_m(\lambda_{N+1};N).
\endaligned\tag 3.17$$
Having obtained $G_{nm}$ as in (3.17) in terms of the
unperturbed quantities related to $V_n(N),$ one can then use
$G_{nm}$ in (2.47) and (2.51) in (2.55) in order to determine
$\varphi_n(\lambda;N+1)$ and $V_n(N+1),$ respectively.

Alternatively, in order to obtain
$\varphi_n(\lambda;N+1)$ and $V_n(N+1),$ we can proceed as follows.
Let us write (3.17) in terms of the real-valued
$(N+1)\times (N+1)$ diagonal matrix $E_N$
and the real-valued column vector $\xi_n$ with $N+1$ entries as
$$G_{nm}=-\ds\frac{C_{N+1}^2}{1-\ds\sum_{k=1}^N C_k^2}\,
\delta_{nm}+\xi_n^\dagger \,E_N\, \xi_m,\tag 3.18$$
where we have defined
$$E_N:=\text{diag}\left\{\ds\frac{C_1^2\, C_{N+1}^2}{1-\ds\sum_{k=1}^N C_k^2},
\ds\frac{C_2^2\, C_{N+1}^2}{1-\ds\sum_{k=1}^N C_k^2},\cdots , \ds\frac{C_N^2\,C_{N+1}^2}{1-\ds\sum_{k=1}^N C_k^2},C_{N+1}^2\right\},\tag 3.19$$
$$\xi_n:=\bm \varphi_n(\lambda_1;N)& \varphi_n(\lambda_2;N)&\cdots
& \varphi_n(\lambda_N;N)&\varphi_n(\lambda_{N+1};N)\endbm^\dagger.\tag 3.20$$
We recall that the dagger in (3.20) can also be replaced
by the matrix transpose since the column vector $\xi_n$ is
real valued. From (3.18) we see that
$G_{nm}$ is separable in $n$ and $m.$ Thus, we can solve the Gel'fand-Levitan system
(2.50) explicitly by seeking $A_{nm}$ in the form
$$A_{nm}=\beta_n^\dagger \,\xi_m,\qquad 1\le m<n,\tag 3.21$$
where the column vector $\beta_n$ has $N+1$ components that are to be determined.
Using (3.18) and (3.21) in (2.50) we observe that $\beta_n^\dagger$ satisfies
$$\beta_n^\dagger+\xi_n^\dagger \,E_N+\beta_n^\dagger\left( -\ds\frac{C_{N+1}^2}{1-\ds\sum_{k=1}^N C_k^2}\,I_{N+1}+
\sum_{j=1}^{n-1} \xi_j\,\xi_j^\dagger \,E_N\right)=0,\tag 3.22$$
where we recall that
$I_{N+1}$ denotes the $(N+1)\times (N+1)$ identity matrix.
 From (3.22) we obtain
$$\beta_n^\dagger=-\xi_n^\dagger E_N
\left(I_{N+1}-\ds\frac{C_{N+1}^2}{1-\ds\sum_{k=1}^N C_k^2}\,I_{N+1}+
\sum_{j=1}^{n-1} \xi_j\,\xi_j^\dagger \,E_N
\right)^{-1},\qquad n=2,3,\dots,\tag 3.23$$
which simplifies to
$$\beta_n^\dagger=-\xi_n^\dagger
\left(\ds\frac{1-\ds\sum_{s=1}^{N+1} C_s^2}{1-\ds\sum_{k=1}^N C_k^2}\,E_N^{-1}+
\sum_{j=1}^{n-1} \xi_j\,\xi_j^\dagger
\right)^{-1},\qquad n=2,3,\dots.\tag 3.24$$
 From (3.21) and (3.24) we see that
$$A_{nm}=- \xi_n^\dagger
\left(\ds\frac{1-\ds\sum_{s=1}^{N+1} C_s^2}{1-\ds\sum_{k=1}^N C_k^2}\,E_N^{-1}+
\ds\sum_{j=1}^{n-1} \xi_j\,\xi_j^\dagger
\right)^{-1}\xi_m,\qquad 1\le m<n.\tag 3.25$$
Hence, for $n\ge 2,$ from (2.51) and (3.25) we obtain
the Darboux transformation at the potential level as
$$\aligned
V_n(N+1)-V_n(N)=&\xi_n^\dagger
\left(\ds\frac{1-\ds\sum_{s=1}^{N+1} C_s^2}{1-\ds\sum_{k=1}^N C_k^2}\,E_N^{-1}+
\ds\sum_{j=1}^{n-1} \xi_j\,\xi_j^\dagger
\right)^{-1}\xi_{n-1}\\
&-\xi_{n+1}^\dagger
\left(\ds\frac{1-\ds\sum_{s=1}^{N+1} C_s^2}{1-\ds\sum_{k=1}^N C_k^2}\,E_N^{-1}+
\ds\sum_{j=1}^n \xi_j\,\xi_j^\dagger
\right)^{-1}\xi_n.\endaligned\tag 3.26$$
Since $A_{10}=0,$ for $n=1,$ instead of (3.26) we need to use
$$V_1(N+1)-V_1(N)=-\xi_2^\dagger
\left(\ds\frac{1-\ds\sum_{s=1}^{N+1} C_s^2}{1-\ds\sum_{k=1}^N C_k^2}\,E_N^{-1}+
\xi_1\,\xi_1^\dagger
\right)^{-1}\xi_1,\tag 3.27$$
which is obtained from (3.26) by replacing the first term
on the right-hand side by zero and by using $n=1$ in the second term.
Note that $\xi_1\,\xi_1^\dagger$ appearing in (3.27) is the $(N+1)\times (N+1)$ matrix
with all entries being equal to one.

Let us remark that (3.25)-(3.27) contain some binomial forms for
the inverse of a matrix. Using (15) on p. 15 of [5], such binomial
forms can be expressed as a ratio of
two determinants. For example, we can write the right-hand side of (3.25) as
$$A_{nm}=\ds\frac{\text{num}}{\text{den}},\tag 3.28$$
where we have defined $\text{num}$ as the determinant of the
$(N+2)\times (N+2)$ block matrix given by
$$ \text{num}:=\det\bm 0& \xi_n^\dagger\\
\stretch
\xi_m &\left(\ds\frac{1-\ds\sum_{s=1}^{N+1} C_s^2}{1-\ds\sum_{k=1}^N C_k^2}\,E_N^{-1}+
\ds\sum_{j=1}^{n-1} \xi_j\,\xi_j^\dagger\right)\endbm,\tag 3.29$$
and we have defined $\text{den}$ as the determinant of the
$(N+1)\times (N+1)$ matrix given by
$$ \text{den}:=\det\left[\ds\frac{1-\sum_{s=1}^{N+1} C_s^2}{1-\ds\sum_{k=1}^N C_k^2}\,E_N^{-1}+
\ds\sum_{j=1}^{n-1} \xi_j\,\xi_j^\dagger\right].\tag 3.30$$

The following theorem shows that the matrix inverses appearing in (3.23)-(3.27)
are well defined and hence the Darboux transformation formulas at the potential
level given in (3.26) and (3.27) are valid.

\noindent {\bf Theorem 3.1} {\it Assume that the potential
$V_n$ appearing in (1.1) belongs to the Faddeev class and that
the discrete Schr\"odinger operator
associated with (1.1) and (1.3) has $N$ bound states with the
corresponding
Gel'fand-Levitan norming constants
$C_s$ defined in (2.39) for $s=1,\dots,N.$ Assume that
an additional bound state is added at $\lambda=\lambda_{N+1}$ with the
Gel'fand-Levitan norming constants
$C_{N+1}.$ Furthermore, assume that
$\sum_{s=1}^{N+1} C_s^2<1.$ Then,
the matrix inverse appearing in (3.25) exists for any $n\ge 2.$}

\noindent PROOF: From (3.19) we see that
$E_N$ is a diagonal matrix with positive entries, and hence
$E_N^{-1}$ is also a diagonal matrix with positive entries.
Then, from (3.25) we see that the matrix whose inverse needs to be established is 
 given by the sum of a diagonal matrix with positive entries and 
 the matrix $\sum_{j=1}^{n-1}\xi\xi^\dagger.$
Let us now consider the
 hermitian form for that sum with any nonzero vector
 $v\in\bC^{N+1}.$ Because the first matrix in the summation 
 is diagonal with positive entries, the corresponding hermitian form
 is strictly positive.
The following argument shows that the hermitian form
 for the second matrix in the summation is nonnegative.
This is established by using
$$v^\dagger \ds\sum_{j=1}^{n-1}\xi_j \, \xi_j^\dagger \,v=\ds\sum_{j=1}^{n-1}\left(\xi_j^\dagger v\right)^\dagger \left(\xi_j^\dagger \,v\right)
=\ds\sum_{j=1}^{n-1}\left|\xi_j^\dagger v\right|^2,\tag 3.31$$
which shows that the right-hand side must be nonnegative.
Thus, the hermitian form with any nonzero vector
$v\in\bC^{N+1}$
associated with the matrix whose inverse is used in (3.25) is positive,
which proves that the matrix itself is positive definite and hence
is invertible. Thus, the right-hand side in (3.25) is well defined when
$\sum_{s=1}^{N+1} C_s^2<1.$ \qed

Let us remark that the case $\sum_{s=1}^{N+1} C_s^2=1$
cannot happen, and hence it is not considered in Theorem~3.1.
This can be seen as follows. If we had $\sum_{s=1}^{N+1} C_s^2=1,$
then (3.12) would imply that $d\rho(\lambda;N+1)=0$ for $\lambda\in[0,4]$
and hence the corresponding discrete Schr\"odinger operator, which
is a selfadjoint operator, would only 
have a discrete spectrum consisting of a finite number of
eigenvalues and no continuous spectrum.
The absence of generalized eigenfunctions as a result of the
absence of the continuous spectrum and the presence of only a finite
number of eigenfunctions related to the discrete spectrum
would be incompatible for the selfadjoint discrete Schr\"odinger operator.
 From the spectral theory we know that the eigenfunctions
 and the generalized eigenfunctions must form a complete set
 acting as an orthogonal basis for the infinite-dimensional
 space of square-summable functions on the half-line lattice,
 and this cannot be done by using only a finite number of
 eigenfunctions.

Let us now evaluate the Darboux transformation
for the regular solution. Using (3.21) in (2.47) we get
$$\varphi_n(\lambda;N+1)=\cases \varphi_n(\lambda;N),\qquad n=1,\\
\varphi_n(\lambda;N)+\beta_n^\dagger\ds\sum_{m=1}^{n-1}
\xi_m  \,\varphi_m(\lambda;N),\qquad n=2,3,\dots.\endcases\tag 3.32$$
As the next proposition shows, the summation term in (3.32) can be written as a linear combination
of $\varphi_{n-1}(\lambda;N)$ and $\varphi_n(\lambda;N).$
Let us define the real-valued column
vector $\alpha_n(\lambda)$ with $N+1$ components as
$$\alpha_n(\lambda):=\bm \ds\frac{\varphi_n(\lambda_1;N)}{\lambda-\lambda_1}& \ds\frac{\varphi_n(\lambda_2;N)}{\lambda-\lambda_2}&\cdots
& \ds\frac{\varphi_n(\lambda_N;N)}{\lambda-\lambda_N}&
\ds\frac{\varphi_n(\lambda_{N+1};N)}{\lambda-\lambda_{N+1}}\endbm^\dagger
,\qquad n\ge 1.\tag 3.33$$

\noindent {\bf Proposition 3.2} {\it Assume that the potential $V_n,$
also denoted by $V_n(N),$
appearing in (1.1)
belongs to the Faddeev class
and the discrete Schr\"odinger operator corresponding
to (1.1) and (1.3) has $N$ bound states at $\lambda=\lambda_s$
with $s=1,\dots,N.$ Let $\varphi_n,$ also denoted by $\varphi_n(\lambda;N),$ be the corresponding
regular solution appearing in (2.4). Let $\xi_n$ be the
real-valued column vector in (3.20) with
$N+1$ components. We then have the following:}

\item{(a)} {\it The summation term in (3.32) can be
simplified and we have}
$$\ds\sum_{m=1}^{n-1}
\xi_m  \,\varphi_m(\lambda;N)=\alpha_n(\lambda)\,\varphi_{n-1}(\lambda;N)-
\alpha_{n-1}(\lambda)\,\varphi_n(\lambda;N),\qquad n=2,3,\dots ,
\tag 3.34$$
{\it where $\alpha_n(\lambda)$ is the
real-valued column vector defined in (3.33)
with $N+1$ components.}

\item{(b)} {\it The $(N+1)\times (N+1)$ matrix consisting of the summation term in (3.24)
can be simplified and its $(k,l)$-component for $n\ge 2$ is given by}
$$\left(\sum_{j=1}^{n-1} \xi_j \,\xi_j^\dagger\right)_{kl}=\cases
\ds\frac{\varphi_{n-1}(\lambda_k;N)\,\varphi_n(\lambda_l;N)-
\varphi_n(\lambda_k;N)\,\varphi_{n-1}(\lambda_l;N)}
{\lambda_k-\lambda_l},
\qquad k\ne l,\\
\stretch
\varphi_n(\lambda_k;N)\,\dot\varphi_{n-1}(\lambda_k;N)-
\varphi_{n-1}(\lambda_k;N)\,\dot\varphi_n(\lambda_k;N),
\qquad k=l,\endcases
\tag 3.35$$
{\it where the overdot denotes the $\lambda$-derivative.}

\noindent PROOF:
Since $\varphi_n(\lambda;N)$ satisfies (1.1) we have
$$\varphi_{m+1}(\lambda;N)+\varphi_{m-1}(\lambda;N)=\left(2+V_m-\lambda\right)\, \varphi_m(\lambda;N),\qquad m=1,2,3,\dots,\tag 3.36$$
$$\varphi_{m+1}(\lambda_s;N)+\varphi_{m-1}(\lambda_s;N)=\left(2+V_m-\lambda_s\right)\, \varphi_m(\lambda_s;N),\qquad m=1,2,3,\dots.\tag 3.37$$
Let us multiply (3.36) by $-\varphi_m(\lambda_s;N)$ and add
(3.37) by $\varphi_m(\lambda;N)$ and add the resulting equations
and then apply the summation over $m$ from $m=1$ to $m=n-1.$ After some
simplifications and using
the first equality
in (2.4), we get
$$\aligned
\varphi_n(\lambda_s;N)\,\varphi_{n-1}(\lambda;N)-&
\varphi_{n-1}(\lambda_s;N)\,\varphi_n(\lambda;N)\\
\stretch
&=
\left(\lambda-\lambda_s\right)\sum_{m=1}^{n-1}\varphi_m(\lambda_s;N)\,\varphi_m(\lambda;N),
\endaligned$$
or equivalently
$$\sum_{m=1}^{n-1}\varphi_m(\lambda_s;N)\,\varphi_m(\lambda;N)=\ds\frac{
\varphi_n(\lambda_s;N)}{\lambda-\lambda_s}\,\varphi_{n-1}(\lambda;N)-\ds\frac{
\varphi_{n-1}(\lambda_s;N)}{\lambda-\lambda_s}\,\varphi_n(\lambda;N).
\tag 3.38$$
Note that (3.38) corresponds to the $s$th component of the
vector relation given in (3.34). Thus, the proof of (a) is complete.
Let us now turn the proof of (b). From (3.20) and the fact that
$\xi_j$ is real, we see that the
$(k,l)$-component of $\xi_j \xi_j^\dagger$ is given by
$$\left(\xi_j \xi_j^\dagger\right)_{kl}=\varphi_j(\lambda_k;N)\,\varphi_j(\lambda_l;N).
\tag 3.39$$
 From (3.38) and (3.39) we see that, when $k\ne l,$ we have
$$\left(\sum_{m=1}^{n-1} \xi_m \,\xi_m^\dagger\right)_{kl}= \ds\frac{
\varphi_n(\lambda_k;N)}{\lambda_l-\lambda_k}\,\varphi_{n-1}(\lambda_l;N)-\ds\frac{
\varphi_{n-1}(\lambda_k;N)}{\lambda_l-\lambda_k}\,\varphi_n(\lambda_l;N),\qquad k\ne l,$$
yielding the first line of (3.35).
When $k=l,$ we can use the limit $\lambda\to\lambda_s$ in (3.38), which gives us
$$\sum_{m=1}^{n-1}\varphi_m(\lambda_s;N)\,\varphi_m(\lambda_s;N)=
\varphi_n(\lambda_s;N)\,\dot\varphi_{n-1}(\lambda_s;N)-
\varphi_{n-1}(\lambda_s;N)\,\dot\varphi_n(\lambda_s;N),$$
yielding the second line of (3.35).
\qed

Using (3.34) in (3.32)
we obtain the Darboux transformation for the regular solution as
$$\varphi_n(\lambda;N+1)=\cases\varphi_n(\lambda;N),\qquad n=1,\\
\stretch
\left[1-\beta_n^\dagger\,\alpha_{n-1}(\lambda)\right]
\varphi_n(\lambda;N)+\beta_n^\dagger\,\alpha_n(\lambda)\,
\varphi_{n-1}(\lambda;N),\qquad n=2,3,\dots,\endcases\tag 3.40$$
where we recall that $\beta_n^\dagger$ is the real-valued row
vector in (3.24), $\alpha_n(\lambda)$
is the real-valued column vector given in (3.33),
and $\xi_n$ is the real-valued column vector given in (3.20).

Note that the results presented in this section remain valid when $N=0.$ In that
case we interpret the summation
$\sum_{k=1}^N C_k^2$ as zero in all the relevant formulas in this section.

\vskip 8 pt
\noindent {\bf 4. DARBOUX TRANSFORMATION IN REMOVING A BOUND STATE}
\vskip 3 pt

In this section we determine the effect of removing a bound state
 from the discrete spectrum of the Schr\"odinger operator
corresponding to (1.1) and (1.3). For clarity, we use the notation
introduced in Section~3. We have the unperturbed potential
$V_n(N)$ containing $N$ bound states
at $\lambda=\lambda_s$ for $s=1,\dots,N.$ We then remove the bound state at
$\lambda=\lambda_N$ with the Gel'fand-Levitan norming constant $C_N$
in order to obtain the perturbed
potential $V_n(N-1)$ containing $N-1$ bound states.
As in Section~3, we
know from (2.38) that there is a one-to-one correspondence between $\lambda_s$ and
$z_s,$ and hence we can equivalently say that the bound states of
the potential $V_n(N)$ occur at $z=z_s$
for $s=1,\dots,N,$ and we remove the bound state at $z=z_N.$

The Darboux transformation for the Jost solution in going from
$f_0(z;N)$ to $f_0(z;N-1)$ can be obtained via (3.6) as
$$f_0(z;N-1)=\ds\frac{1-z_N z}{1-\ds\frac{z}{z_N}}\,f_0(z;N),\qquad |z|\le 1.
\tag 4.1$$
Similarly, the Darboux transformation for the scattering matrix in going from
$S(z;N)$ to $S(z;N-1)$ can be obtained via (3.8) as
$$S(z;N-1)=\left(\ds\frac{z-z_N}{1-z_N z}\right)^2\,S(z;N),\qquad z\in\bold T.$$
With the help of (3.9) we see that the Darboux transformation for the
phase shift in going from $\phi(z;N)$ to $\phi(z;N-1)$ can be obtained
via (3.9) as
$$\phi(z;N-1)=\phi(z;N)+\ds\frac{i}{2}\,\log\left(\ds\frac{1-z_N\,z}{z-z_N}\right)^2,
\qquad z\in\bold T.$$

Let us now determine the Darboux transformation for the spectral density
in going from $d\rho(\lambda;N)$ to $d\rho(\lambda;N-1).$ From (3.10)
we see that
$$d\rho(\lambda;N-1)=\cases \ds\frac{1-\ds\sum_{s=1}^{N-1} C_s^2}{
\ds\prod_{k=1}^{N-1} z_k^2}\,\ds\frac{ d\overset{\circ}\to\rho}{|f_0(z;N-1)|^2},
\qquad \lambda\in[0,4],\\
\stretch
\ds\sum_{s=1}^{N-1} C_s^2\,\delta(\lambda-\lambda_s)\,d\lambda,\qquad
\lambda\in\bR\setminus[0,4].\endcases\tag 4.2$$
On the other hand, from (3.5) we have
$$|f_0(z;N-1)|^2=z_N^2\, |f_0(z;N)|^2,\qquad z\in\bold T.\tag 4.3$$
Using (4.3) in (4.2) we get
$$d\rho(\lambda;N-1)=\cases \ds\frac{1-\ds\sum_{s=1}^{N-1} C_s^2}{
\ds\prod_{k=1}^N z_k^2}\,\ds\frac{ d\overset{\circ}\to\rho}{|f_0(z;N)|^2},
\qquad \lambda\in[0,4],\\
\stretch
\ds\sum_{s=1}^{N-1} C_s^2\,\delta(\lambda-\lambda_s)\,d\lambda,\qquad
\lambda\in\bR\setminus[0,4].\endcases\tag 4.4$$
We recall that $\lambda\in[0,4]$ in (4.2) and (4.4) corresponds to
 $z\in \overline{\bold T^+}.$
Thus, from (3.10) and (4.4) we get
$$ d\rho(\lambda;N-1)-d\rho(\lambda;N)=
\cases \ds\frac{C_N^2}{1-\ds\sum_{s=1}^N C_s^2}
\,d\rho(\lambda;N),
\qquad \lambda\in[0,4],\\
\stretch
-C_N^2\,\delta(\lambda-\lambda_N)\,d\lambda,\qquad
\lambda\in\bR\setminus[0,4].\endcases\tag 4.5$$

Next, we determine the Darboux transformation for the
regular solution in going from $\varphi_n(\lambda;N)$ to
$\varphi_n(\lambda;N-1).$
In the Gel'fand-Levitan formalism outlined in (2.47)-(2.51), we
have
$$\varphi_n(\lambda;N-1)=\cases
\varphi_n(\lambda;N),\qquad n=1,\\
\stretch
\varphi_n(\lambda;N)+\ds\sum_{m=1}^{n-1}
A_{nm}\,\varphi_m(\lambda;N),\qquad n=2,3,\dots ,
\endcases$$
$$G_{nm}:=\int_{\lambda\in\bR} \varphi_n(\lambda;N)\left[
d\rho(\lambda;N-1)-d\rho(\lambda;N)\right]\varphi_m(\lambda;N),\tag 4.6$$
where the constants $A_{nm}$ are to be determined from (2.50)
by using (4.6) as input. In this case, from (2.51) we get
$$V_n(N-1)-V_n(N)=A_{(n+1)n}-A_{n(n-1)},\qquad n=1,2,3,\dots ,$$
again with the understanding that $A_{10}=0.$
Using (4.5) in (4.6) we obtain
$$\aligned
G_{nm}=&\ds\frac{C_N^2}{1-\ds\sum_{k=1}^N C_k^2}
\,\int_{\lambda\in [0,4]} \varphi_n(\lambda;N)\,d\rho(\lambda;N)\, \varphi_m(\lambda;N)\\
\stretch
&-C_N^2 \,\varphi_n(\lambda_N;N)\,\varphi_m(\lambda_N;N).
\endaligned\tag 4.7$$
Using (3.15) in (4.7), after some simplification we get
$$\aligned
G_{nm}=&\ds\frac{C_N^2}{1-\ds\sum_{k=1}^N C_k^2}\,
\delta_{nm}-\ds\frac{C_N^2}{1-\ds\sum_{k=1}^N C_k^2}
\,\ds\sum_{s=1}^{N-1} C_s^2\,
\varphi_n(\lambda_s;N)\,\varphi_m(\lambda_s;N)\\
&-C_N^2 \,\ds\frac{1-\ds\sum_{s=1}^{N-1} C_s^2}{1-\ds\sum_{k=1}^N C_k^2}\,
\varphi_n(\lambda_N;N)\,\varphi_m(\lambda_N;N).
\endaligned\tag 4.8$$
Proceeding as in (3.18)-(3.20) we can write $G_{nm}$ given in
(4.8) as
$$G_{nm}=\ds\frac{C_N^2}{1-\ds\sum_{k=1}^N C_k^2}\,
\delta_{nm}+\theta_n^\dagger \,F_N\, \theta_m,\tag 4.9$$
where $F_N$ is the $N\times N$ diagonal matrix with real entries given by
$$F_N:=\text{diag}\left\{\ds\frac{-C_1^2\, C_N^2}{1-\ds\sum_{k=1}^N C_k^2},
\ds\frac{-C_2^2\, C_N^2}{1-\ds\sum_{k=1}^N C_k^2},\cdots, \ds\frac{-C_{N-1}^2\,C_N^2}{1-\ds\sum_{k=1}^N C_k^2},\ds\frac{-C_N^2\left( 1-\ds\sum_{s=1}^{N-1} C_s^2 \right)}{1-\ds\sum_{k=1}^N C_k^2}\right\},\tag 4.10$$
$$\theta_n:=\bm \varphi_n(\lambda_1;N)& \varphi_n(\lambda_2;N)&\cdots
& \varphi_n(\lambda_{N-1};N)&\varphi_n(\lambda_N;N)\endbm^\dagger.\tag 4.11$$
Comparing (3.20) and (4.11) we observe that the first $N$ entries of
the column vectors
$\theta_n$ and $\xi_n$ are identical and that $\xi_n$ has an additional
$(N+1)$st entry.
As in Section~3, the quantity $G_{nm}$ given in (4.9) is separable in
$n$ and $m,$ and hence the Gel'fand-Levitan system (2.50) is explicitly
solvable by using the analog of (3.21), i.e. by letting
$$A_{nm}=\gamma_n^\dagger \theta_m,\qquad 1\le m<n,\tag 4.12$$
where the column vector $\gamma_n$ has $N$ components to be determined.
Proceeding as in (3.22)-(3.25) we determine $\gamma_n^\dagger$ as
$$\gamma_n^\dagger=-\theta_n^\dagger
\left(\ds\frac{1-\ds\sum_{s=1}^{N-1} C_s^2}{1-\ds\sum_{k=1}^N C_k^2}\,F_N^{-1}+
\ds\sum_{j=1}^{n-1} \theta_j\,\theta_j^\dagger
\right)^{-1}.\tag 4.13$$
 From (4.12) and (4.13) we see that
$$A_{nm}=- \theta_n^\dagger
\left(\ds\frac{1-\ds\sum_{s=1}^{N-1} C_s^2}{1-\ds\sum_{k=1}^N C_k^2}\,F_N^{-1}+
\ds\sum_{j=1}^{n-1} \theta_j\,\theta_j^\dagger
\right)^{-1}\theta_m,\qquad 1\le m<n.\tag 4.14$$

The analogs of (3.28)-(3.30) also apply in this case. Since the right-hand side of (4.12) is a binomial for a matrix inverse, we can write
$A_{nm}$ given in (4.12) as the ratio of two determinants as
$$A_{nm}=\ds\frac{\det\bm 0& \theta_n^\dagger\\
\stretch
\theta_m &\left(\ds\frac{1-\ds\sum_{s=1}^{N-1} C_s^2}{1-\ds\sum_{k=1}^N C_k^2}\,F_N^{-1}+
\sum_{j=1}^{n-1} \theta_j\,\theta_j^\dagger
\right)\endbm}{\det\left[\ds\frac{1-\ds\sum_{s=1}^{N-1} C_s^2}{1-\ds\sum_{k=1}^N C_k^2}\,F_N^{-1}+
\ds\sum_{j=1}^{n-1} \theta_j\,\theta_j^\dagger\right]},\qquad 1\le m<m.\tag 4.15$$

As in Proposition~3.2(b), for $n\ge 2$ we can simplify the
$N\times N$ matrix $\sum_{j=1}^{n-1}\theta_j \theta_j^\dagger$
appearing in (4.13)-(4.15) and find that its $(k,l)$-entry
is given by
$$\left(\sum_{j=1}^{n-1} \theta_j \,\theta_j^\dagger\right)_{kl}=\cases
\ds\frac{\varphi_{n-1}(\lambda_k;N)\,\varphi_n(\lambda_l;N)-
\varphi_n(\lambda_k;N)\,\varphi_{n-1}(\lambda_l;N)}
{\lambda_k-\lambda_l},
\qquad k\ne l,\\
\stretch
\varphi_n(\lambda_k;N)\,\dot\varphi_{n-1}(\lambda_k;N)-
\varphi_{n-1}(\lambda_k;N)\,\dot\varphi_n(\lambda_k;N),
\qquad k=l.\endcases
\tag 4.16$$
Let us remark that the matrix in (3.35) has $N+1$
rows and $N+1$
columns, and the matrix in (4.16) has $N$ rows and $N$ columns.
If we delete the $(N+1)$st row and $(N+1)$st column from the matrix
in (3.35) we get the matrix
in (4.16).

The analog of (3.26) in this case is obtained by using (4.14) in (2.51),
and for $n\ge 2$
we get the Darboux transformation in going from $V_n(N)$ to $V_n(N-1)$
given by
$$\aligned
V_n(N-1)-V_n(N)=&\theta_n^\dagger
\left(\ds\frac{1-\ds\sum_{s=1}^{N-1} C_s^2}{1-\ds\sum_{k=1}^N C_k^2}\,F_N^{-1}+
\ds\sum_{j=1}^{n-1} \theta_j\,\theta_j^\dagger
\right)^{-1}\theta_{n-1}\\
&-\theta_{n+1}^\dagger
\left(\ds\frac{1-\ds\sum_{s=1}^{N-1} C_s^2}{1-\ds\sum_{k=1}^N C_k^2}\,F_N^{-1}+
\ds\sum_{j=1}^n \theta_j\,\theta_j^\dagger
\right)^{-1}\theta_n.\endaligned\tag 4.17$$
For $n=1$, instead of (4.17) we use the analog of (3.27) and get
$$V_1(N-1)-V_1(N)=-\theta_2^\dagger
\left(\ds\frac{1-\ds\sum_{s=1}^{N-1} C_s^2}{1-\ds\sum_{k=1}^N C_k^2}\,F_N^{-1}+
\theta_1\,\theta_1^\dagger
\right)^{-1}\theta_1.\tag 4.18$$

The analog of (3.32) in this case is
$$\varphi_n(\lambda;N-1)=\cases \varphi_n(\lambda;N),\qquad n=1,\\
\stretch
\varphi_n(\lambda;N)+\gamma_n^\dagger\ds\sum_{m=1}^{n-1}
\theta_m  \,\varphi_m(\lambda;N),\qquad n=2,3,\dots,\endcases$$
and the analog of (3.40) in this case is
$$\varphi_n(\lambda;N-1)=\cases \varphi_n(\lambda;N),\qquad n=1,\\
\stretch
\left[1-\gamma_n^\dagger\,\epsilon_{n-1}(\lambda)\right]
\varphi_n(\lambda;N)+\gamma_n^\dagger\,\epsilon_n(\lambda)\,
\varphi_{n-1}(\lambda;N),\qquad n=2,3,\dots,\endcases$$
where $\epsilon_n(\lambda)$ is the column vector with $N$ components and
it is defined as
$$\epsilon_n(\lambda):=\bm \ds\frac{\varphi_n(\lambda_1;N)}{\lambda-\lambda_1}& \ds\frac{\varphi_n(\lambda_2;N)}{\lambda-\lambda_2}&\cdots
& \ds\frac{\varphi_n(\lambda_{N-1};N)}{\lambda-\lambda_{N-1}}&
\ds\frac{\varphi_n(\lambda_N;N)}{\lambda-\lambda_N}\endbm^\dagger,\qquad n\ge 1.\tag 4.19$$
We remark that the column vector $\epsilon_n(\lambda)$ given in
(4.19) has $N$ components, and the column vector
$\alpha_n(\lambda)$ given in (3.33) has $N+1$ components.
In fact, $\epsilon_n(\lambda)$ is obtained from
$\alpha_n(\lambda)$ by omitting the last entry.

In the following theorem we present the analog of the result presented
in Theorem~3.1, i.e. we prove that the matrix inverse appearing in
(4.14)is well defined and hence the Darboux
transformation formulas at the potential level given
in (4.17) and (4.18) are valid.
Let us remark that the matrix in (3.25) whose inverse 
is established in Theorem~3.1 consists of the sum of
a diagonal matrix with positive entries and
a nonnegative definite hermitian matrix. In contrast,
the matrix in (4.14) whose inverse is established in the next
theorem consists of the sum of
a diagonal matrix with negative entries and
a nonnegative definite hermitian matrix.

\noindent {\bf Theorem 4.1} {\it Assume that the potential
$V_n$ appearing in (1.1) belongs to the Faddeev class and that
the discrete Schr\"odinger operator
associated with (1.1) and (1.3) has $N$ bound states with the
corresponding
Gel'fand-Levitan norming constants
$C_s$ defined in (2.39) for $s=1,\dots,N.$ Assume that
the bound state at $\lambda=\lambda_N$ with the
Gel'fand-Levitan norming constants
$C_N$ is removed from the discrete spectrum. Furthermore, assume that
$\sum_{s=1}^N C_s^2<1.$ Then,
the matrix inverse appearing in (4.14) exists for any $n\ge 2.$}

\noindent PROOF: As a result of the assumption
$\sum_{s=1}^N C_s^2<1,$ from (4.10) we observe that
each entry of the diagonal matrix $F_N$ given in
(4.10) is negative and hence
$F_N^{-1}$ is also a diagonal matrix with negative
entries. We can write
the matrix in (4.14) whose inverse is to be established
as $-H_N+\sum_{j=1}^{n-1} \theta_j\,\theta_j^\dagger,$ where
we have defined
$$H_N:=-\ds\frac{1-\ds\sum_{s=1}^{N-1} C_s^2}{1-\ds\sum_{k=1}^N C_k^2}\,F_N^{-1}.
\tag 4.20$$
Using (4.10) in (4.20) we obtain
$$H_N=\ds\frac{1-\ds\sum_{s=1}^{N-1} C_s^2}{C_N^2}\,
\text{diag}\left\{\ds\frac{1}{C_1^2},\ds\frac{1}{C_2^2},
\cdots,\ds\frac{1}{C_{N-1}^2},\ds\frac{1}{1-\ds\sum_{k=1}^{N-1} C_k^2}\right\}.\tag 4.21$$
We let
$$\varepsilon_N:=\ds\frac{1-\ds\sum_{s=1}^N C_s^2}{C_N^2}.\tag 4.22$$
and observe that $\varepsilon_N$ is a positive number as a result of
$\sum_{s=1}^N C_s^2<1.$
Note that
$$\ds\frac{1-\ds\sum_{k=1}^{N-1} C_k^2}{C_N^2}=\ds\frac{C_N^2+1-\ds\sum_{k=1}^N C_k^2}{C_N^2}=1+\ds\frac{1-\ds\sum_{s=1}^N C_s^2}{C_N^2}.\tag 4.23$$
With the help of (4.22) and (4.23) we write (4.21) as
$$H_N=
\text{diag}\left\{\ds\frac{1+\varepsilon_N}{C_1^2},\ds\frac{1+\varepsilon_N}{C_2^2},
\cdots,\ds\frac{1+\varepsilon_N}{C_{N-1}^2},\ds\frac{1}{C_N^2}\right\}.\tag 4.24$$
Let $v$ be a nonzero vector in $\bC^N$ given by
$$v=\bm v_1\\
\vdots\\
v_{N+1}\endbm,\tag 4.25$$
The hermitian form of $H_N$ with the vector $v$ given in (4.25) is obtained from (4.23) as
$$v^\dagger H_N \,v=\ds\frac{(1+\varepsilon_N)\,|v_1|^2}{C_1^2}+\ds\frac{(1+\varepsilon_N)\,|v_2|^2}{C_2^2}+
\cdots+\ds\frac{(1+\varepsilon_N)\,|v_{N-1}|^2}{C_{N-1}^2}+\ds\frac{|v_N|^2}{C_N^2}.
\tag 4.26$$
Since $\varepsilon_N>0,$ from (4.26) we obtain
$$v^\dagger H_N \,v\ge \ds\frac{|v_1|^2}{C_1^2}+\ds\frac{|v_2|^2}{C_2^2}+
\cdots+\ds\frac{|v_{N-1}|^2}{C_{N-1}^2}+\ds\frac{|v_N|^2}{C_N^2}.
\tag 4.27$$
We evaluate the hermitian form of $\sum_{j=1}^{n-1} \theta_j\,\theta_j^\dagger$ with
the vector $v$ given in (4.25)
as in (3.31) and obtain
$$v^\dagger \ds\sum_{j=1}^{n-1} \theta_j\,\theta_j^\dagger v=
\ds\sum_{j=1}^{n-1}\left|\theta_j^\dagger v\right|^2,
\tag 4.28$$
 from (4.28) we conclude that
$$v^\dagger \ds\sum_{j=1}^{n-1} \theta_j\,\theta_j^\dagger v<
\ds\sum_{j=1}^\infty\left|\theta_j^\dagger v\right|^2,
\tag 4.29$$ 
where we have used the fact that we cannot have
$\theta_j^\dagger v=0$ for all $j\ge n.$
Using (4.11) and (4.25) we get
$$\theta_j^\dagger\, v=\varphi_j(\lambda_1;N)\,v_1+\varphi_j(\lambda_2;N)\,v_2+
\cdots+\varphi_j(\lambda_N;N)\,v_N,\tag 4.30$$
where we recall that each entry in (4.11) is real. From (4.30) we obtain
$$\left|\theta_j^\dagger v\right|^2=\ds\sum_{k=1}^N
\varphi_j(\lambda_k;N)^2\,|v_k|^2+
2\ds\sum_{1=k<l\le N}\varphi_j(\lambda_k;N)\,\varphi_j(\lambda_l;N)\,v_k^\ast\,v_l.
\tag 4.31$$
Since the discrete Schr\"odinger operator
associated with (1.1) and (1.3) is selfadjoint, eigenvectors
corresponding to distinct eigenvalues are orthogonal and we have
$$\ds\sum_{j=1}^\infty \varphi_j(\lambda_k;N)\,\varphi_j(\lambda_l;N)=0,
\qquad k\ne l.\tag 4.32$$
Thus, with the help of (4.32), from (4.31) we get
$$\ds\sum_{j=1}^\infty\left|\theta_j^\dagger v\right|^2=
\ds\sum_{k=1}^N\left(\ds\sum_{j=1}^\infty
\varphi_j(\lambda_k;N)^2\right)\,|v_k|^2.\tag 4.33$$
Using (2.39) in (4.33) we get
$$\ds\sum_{j=1}^\infty\left|\theta_j^\dagger v\right|^2=
\ds\sum_{k=1}^N\ds\frac{|v_k|^2}{C_k^2}
.\tag 4.34$$
Thus, from (4.29) and (4.34) we get
$$v^\dagger \ds\sum_{j=1}^{n-1} \theta_j\,\theta_j^\dagger v<
\ds\frac{|v_1|^2}{C_1^2}+\ds\frac{|v_2|^2}{C_2^2}+\cdots+\ds\frac{|v_N|^2}{C_N^2}.
\tag 4.35$$ 
Combining (4.27) and (4.35) we obtain
$$v^\dagger \left(-H_N+\ds\sum_{j=1}^{n-1}\theta_j\,\theta_j^\dagger\right) v<0.\tag 4.36$$
 From (4.36) we conclude that the matrix
 whose inverse appears in (4.14) is negative definite and hence
 that matrix must be invertible. \qed


\vskip 8 pt
\noindent {\bf 5. SOME EXPLICIT EXAMPLES}
\vskip 3 pt

In this section we illustrate the results of the previous sections with some
explicit examples. We also make some contrasts between the Darboux
transformation for (1.1) and the Darboux
transformation for (1.2) when the potentials are compactly supported.

Let us consider the case where the potential $V_n$ in (1.1) is nontrivial
and compactly supported, i.e.
assume that $V_n=0$ for $n>b$ and $V_b\ne 0$ for some positive integer $b.$
The corresponding Jost function $f_0$ appearing in (2.10) is then
a polynomial in $z$ of degree $2b-1$ and, as (2.50) of [2] indicates,
is given by
$$f_0=1+z\ds\sum_{j=1}^b V_j+\cdots+z^{2b-2}\ds\sum_{j=1}^{b-1} V_b\, V_j+
z^{2b-1} V_b.\tag 5.1$$
For a compactly-supported potential,
 the Marchenko norming constant
 $c_s$ defined in
 (2.40) is obtained [2] from the residue
 of $S/z$
 at the bound-state value $z_s$
 as
$$c_s^2=\text{Res}\left[\ds\frac{S}{z},z_s\right],\qquad s=1,\dots,N,
\tag 5.2$$
where $S$ is the scattering matrix defined in (2.10). Then, the
corresponding Gel'fand-Levitan norming constant $C_s$ can be obtained
by using (2.42).

In some of the examples in this section, we illustrate that
not every polynomial in
$z$ of degree $2b-1$ necessarily corresponds to the
Jost function $f_0$ of a compactly-supported potential vanishing
for $n>b.$ This is not surprising because the coefficients in such a polynomial
must agree with the coefficients given in (5.1). There are
$b$ potential values that need to correspond to the $(2b-1)$
coefficients
on the right-hand side of (5.1). For example, when
$b=2$ from (5.1) we get
$$f_0=1+(V_1+V_2)z+V_1 V_2 z^2+V_2 z^3,\tag 5.3$$
and the same quantity must also have the form
$$f_0=\left(1-\ds\frac{z}{\alpha_1}\right)\left(1-\ds\frac{z}
{\alpha_2}\right)\left(1-\ds\frac{z}{\alpha_3}\right),
\tag 5.4$$
for some nonzero constants $\alpha_1,$ $\alpha_2,$ $\alpha_3$
satisfying
$$\cases
V_1+V_2=-\left(\ds\frac{1}{\alpha_1}+\ds\frac{1}{\alpha_2}+\ds\frac{1}{\alpha_3}\right),
\\
V_1 V_2=\ds\frac{1}{\alpha_1\,\alpha_2}+\ds\frac{1}{\alpha_1\,\alpha_3}+\ds\frac{1}{\alpha_2\,\alpha_3},\\
V_2=-\ds\frac{1}{\alpha_1\,\alpha_2\,\alpha_3}.\endcases\tag 5.5$$
In case the system (5.5) is inconsistent, the quantity
given on the right-hand side of (5.4) cannot be the Jost solution of a compactly-supported
potential.

For the half-line Schr\"odinger equation (1.2) with a compactly-supported
potential $V(x),$ the following property is known [1]. If we remove a bound state
 from such a potential, then the transformed potential is also
 compactly supported and the transformed potential is guaranteed to vanish outside
 the support of the original potential.
In some of the examples in this section, we illustrate that
 the aforementioned support property
does not necessarily hold for the discrete Schr\"odinger equation (1.1).
We show that the property holds in one example but does not hold
 in another example.

For the half-line Schr\"odinger equation (1.2) with a compactly-supported
potential $V(x),$ also the following second property holds [1].
If we add a bound state
to a compactly-supported potential, then the transformed potential is also
 compactly supported (and the transformed potential is guaranteed to vanish outside
 the support of the original potential) if and only if the
 two conditions specified in Theorem~3.5 of [1] are satisfied.
 The first condition is that the added bound-state $\lambda_s$-value
 must come from an ``eligible" resonance [1] and the second condition is
 that the corresponding Gel'fand-Levitan norming constant
 $C_s$ must have a specific positive value.
In some of the examples in this section,
we illustrate that
 the aforementioned support property
does not necessarily hold for the discrete Schr\"odinger equation (1.1).
We show that the property holds in one example but does not hold
 in another example.

In the next example, we add a bound state at
$z=z_1$ with the Gel'fand-Levitan norming constant $C_1$
 to a compactly-supported potential with $b=1.$ The example shows
 that the Darboux transformation on
the compactly-supported potential results in a compactly-supported
potential if the values for $z_1$ and $C_1$ are chosen appropriately.

\noindent {\bf Example 5.1} Consider the compactly-supported potential
$V_n$ with $b=1$ and hence $V_n=0$ for $n\ge 2.$ Let us assume that $0<|V_1|\le 1.$
 From (5.1) we see that the
Jost function is given by
$$f_0=1+V_1 z.\tag 5.6$$
Using (2.4) in (2.3), we obtain the corresponding regular solution $\varphi_n$
as a function of $z$ as
$$\varphi_n=\ds\frac{z^n-z^{-n}}{z-z^{-1}}+V_1\,\ds\frac{z^{n-1}-z^{1-n}}{z-z^{-1}},
\qquad n=1,2,\dots.\tag 5.7$$
Since the bound states correspond to the zeros of $f_0$
when $z\in(-1,0)\cup(0,1),$ from (5.6) we see that there are no bound states
and hence we have $N=0.$ Let us now add one bound state at $z=z_1$ with the Gel'fand-Levitan norming constant $C_1.$ Let us choose $z_1=-V_1,$
and hence impose the further restriction $0<|V_1|<1.$
Let us use $\tilde f_0$ and
$\tilde V_n$ to denote the corresponding Jost function and potential, respectively,
when the bound state is added. From (3.6) and (5.6)
we see that
$$\tilde f_0=1+z/V_1.\tag 5.8$$
Using (5.7) and $z_1=-V_1$ in (3.20), we obtain
$$\xi_n=(-V_1)^{1-n},\qquad n=1,2,\dots.$$
The quantity $E_N$ defined in (3.19) with $N=0$ is given by
$E_0=C_1^2.$ Then, (3.27) and (3.26) respectively yield
$$\tilde V_1=V_1+\ds\frac{C_1^2}{V_1},\tag 5.9$$
$$\tilde V_n=\ds\frac{-C_1^2 V_1^{2n+1}(1-V_1^2)^2 (C_1^2-1+V_1^2)}
{C_1^2 V_1^6-C_1^2  V_1^{2n+2} (1+V_1^2)(C_1^2-1+V_1^2)+V_1^{4n} (C_1^2-1+V_1^2)^2},
\qquad n\ge 2.\tag 5.10$$
 From (5.10) we see that $\tilde V_n$ is compactly supported
 if and only if we have
 $$C_1^2=1-V_1^2.\tag 5.11$$
 In fact, with the special choice of
 the Gel'fand-Levitan norming constant in (5.11), from (5.9) we obtain
 $\tilde V_1=1/V_1.$ In the presence of one bound state
 for the compactly-supported potential $\tilde V_n,$
 the corresponding Gel'fand-Levitan norming constant
 $C_1$ can be evaluated with the help of (2.41), (5.2), (5.8),
 and the fact that $\tilde f_1=z,$ yielding the value of
 $C_1^2$ given in (5.11).

In the following example, we illustrate that a polynomial
in $z$ of degree $2b-1$ may or may not correspond to the Jost function of
a compactly-supported potential.

\noindent {\bf Example 5.2} Consider the Jost function $f_0$ given by
$$f_0=(1+2z)(1-2z)\left(1-\ds\frac{z}{\sqrt{5}}\right).$$
Comparing (5.2) with (5.1), we see that one solution to
the corresponding system (5.2) results in
$$b=2,\quad V_1=-\sqrt{5},\quad V_2=\ds\frac{4}{\sqrt{5}}.\tag 5.12$$
 From (5.2) we see that $f_0$ has two zeros when
 $z\in(-1,0)\cup(0,1),$ and hence it has two
 bound-state zeros
given by
 $z_1=-1/2$ and $z_2=1/2.$
  From (2.46) we see that the corresponding
  Gel'fand-Levitan norming constants
  $C_1$ and $C_2$ must satisfy
 $0< C_1^2+C_2^2\le 1.$
Corresponding to a compactly-supported potential we must [2] have
$f_n=z^n$ for $n\ge b.$ Hence, in our example, corresponding to (5.6) we have
$f_2=z^2$ and $f_3=z^3.$ Then, from (2.3) with $n=2$
we obtain $f_1(z)=z+V_2 z^2.$ With the help of (2.41), (2.42), and (5.2), we get
$$C_1^2=\ds\frac{3(12-5\sqrt{5})}{76}=0.03235\overline{5}
  ,\quad C_2^2= \ds\frac{3(12+5\sqrt{5})}{76}=0.91501\overline{3},\tag 5.13$$
  where the overline on a digit indicates a round off.
  We note that (5.13) is compatible with the constraint $0< C_1^2+C_2^2\le 1.$
Thus, we have confirmed that $z_1=-1/2$ and $z_2=1/2$
do indeed correspond to bound states of the compactly-supported potential
described in (5.12).
  In (5.4), if we choose $\alpha_j=1$ for $j=1,2,3,$ then the system in (5.5)
  becomes inconsistent and hence there are no values $V_1$ and $V_2$ satisfying
  (5.5). Thus, the corresponding expression in (5.4)
  does not yield a compactly-supported potential.
  On the other hand, if we let $V_1=-\sqrt{2}$ and $V_2=1/\sqrt{2}$ in (5.3), we
  get a solution to (5.5) with $\alpha_1=-1,$ $\alpha_2=1,$ and $\alpha_3=\sqrt{2},$ and hence
  the Jost solution obtained from (5.4) does not contain any zeros
  in $z\in(-1,0)\cup(0,1),$ yielding $N=0.$ Choosing
  $V_1=-(7+\sqrt{10})/6$ and $V_2=-(1+\sqrt{10})/2$ in (5.3), we get a solution to
  (5.5) given by
$$\alpha_1=\ds\frac{3}{2(1+\sqrt{10})}=0.3603\overline{8},\quad \alpha_2=\ds\frac{2}{1+\sqrt{2} i},\quad \alpha_3=\ds\frac{2}{1-\sqrt{2} i},$$
which indicates that the corresponding $f_0$ in (5.4) has one
bound state at $z_1=\alpha_1$ with the corresponding
Gel'fand-Levitan constant $C_1,$ evaluated
with the help of (2.40), (2.42), and (5.2), as
$$C_1^2=\ds\frac{625+128\sqrt{10}}{3489}=0.29514\overline{8}.$$

We remark that it is impossible to have a compactly-supported
potential with $b=2$ having three bound states. This can be seen as
follows. Assume that for some choice of
$V_1$ and $V_2$ in (5.3) we had $-1<\alpha_1<\alpha_2<\alpha_3<1$ for
nonzero $\alpha_j$ values. Using (5.4) in (2.10) and (5.2) we would get
the corresponding Marchenko norming constants as
$$\cases c_1^2=\ds\frac{(1-\alpha_1^2)(1-\alpha_1 \alpha_2)(1-\alpha_1 \alpha_3)}{\alpha_1^4 (\alpha_2-\alpha_1)(\alpha_3-\alpha_1)},\\ c_2^2=\ds\frac{(1-\alpha_1 \alpha_2)(1-\alpha_2^2 )(1-\alpha_2 \alpha_3)}{\alpha_2^4 (\alpha_1-\alpha_2)(\alpha_3-\alpha_2)},\\ c_3^2=\ds\frac{(1-\alpha_1 \alpha_3)(1-\alpha_2 \alpha_3)(1-\alpha_3^2)}{\alpha_3^4 (\alpha_1-\alpha_3)(\alpha_2-\alpha_3)}.
\endcases \tag 5.14$$
 From (5.14) we see that we would have $c_1^2>0,$ $c_2^2<0,$ $c_3^2>0,$ and hence
it is impossible to have $N=3.$
 From Example~5.1 we know that $0\le N\le b$ when $b=1,$ and
 from (5.14) we know that $0\le N\le b$ when $b=2.$ From (5.1)
 it is clear that the number of zeros of $f_0(z)$ in $z\in(-1,0)\cup(0,1)$
 cannot exceed $2b-1.$ We pose the following as an open problem,
 which can perhaps be answered with the help of a generalization
 of (5.14) from $b=2$ to an arbitrary positive integer $b:$ For any given
 positive integer $b,$ what is the maximal number of bound states for the
 corresponding Schr\"odinger operator associated with (1.1) and (1.4), if
the potential $V_n$ has a compact support with $V_n=0$ for $n>b$?

The regular solution $\varphi_n$ to
(1.1) corresponding to (5.3) can be obtained
recursively with the help of (2.4). We have
$$\varphi_1=1,\quad \varphi_2=-\lambda+2+V_1,
\quad \varphi_3=\lambda^2-(4+V_1+V_2)\lambda+3+2V_1+2V_2+V_1 V_2,
\tag 5.15$$
$$\varphi_4=-\lambda^3+(6+V_1+V_2)\lambda^2-(10+4V_1+4 V_2+V_1 V_2)\lambda+4+3V_1+4V_2+
2 V_1 V_2,
\tag 5.16$$
$$\aligned
\varphi_5=&\lambda^4-(8+V_1+V_2)\lambda^3+(21+6 V_1+6 V_2+V_1 V_2)\lambda^2\\
&-(20+10 V_1+11  V_2+4 V_1 V_2)\lambda+5+4V_1+6 V_2+
3 V_1 V_2.\endaligned
\tag 5.17$$

In the next two examples, we show that if we remove a bound state from
a compactly-supported potential then the resulting potential may or may not
be compactly supported.

\noindent {\bf Example 5.3} Consider the compactly-supported potential
$V_n$ with $b=1$ and hence $V_n=0$ for $n\ge 2.$ The corresponding Jost
function is given by (5.6). Since the bound states correspond to the zeros of $f_0$
when $z\in(-1,0)\cup(0,1),$ from (5.6) we see that there exists one bound state if $|V_1|>1.$ We assume that $|V_1|>1$ so that
we have exactly one bound state at $z=z_1,$ where
$z_1=-1/V_1.$
 From (2.10) and (5.6) we see that the corresponding scattering matrix
 is given by
 $$S(z)=\ds\frac{V_1+z}{z+V_1 z^2},\qquad z\in\bold T.\tag 5.18$$
In this case, the Jost solution satisfies $f_n=z^n$ for $n\ge 1.$
 In the presence of one bound state, the corresponding Gel'fand-Levitan
 norming constant
$C_1$ is evaluated with the help of (2.42), (5.2), (5.18), and $f_1=z,$ yielding
$$C_1^2=V_1^2-1.\tag 5.19$$
 From (2.46) we see that we must have $0<C_1^2\le 1$ and hence
 we must use the restriction $0<|V_1|\le \sqrt{2}.$
Let us now remove the bound state with $z_1=-1/V_1.$ The transformed Jost solution $\tilde f_0$ is
obtained via (4.1) and is given by
$\tilde f_0=1+z/V_1.$ In this case, using (4.11) and (5.7) we obtain
$$\theta_n=\left(-\ds\frac{1}{V_1}\right)^{n-1},\qquad n=1,2,\dots.
\tag 5.20$$
Using (5.19) with $N=1,$ we get the quantity $F_N$ given in
(4.10) as
$$F_1=1-V_1^2.\tag 5.21$$
Using (5.20) and (5.21) in (4.17) and (4.18) we obtain
$\tilde V_n=0$ for $n\ge 2$ and $\tilde V_1=1/V_1.$

\noindent {\bf Example 5.4} Consider the compactly-supported
potential $V_n$ described by (5.12) in Example~5.2. We know
 from Example~5.2 that there are two bound states with
 $z_1=-1/2$ and $z_2=1/2$ with the respective
 corresponding Gel'fand-Levitan norming constants
 $C_1$ and $C_2$ as in (5.13).
Hence, we have $N=2.$ We now demonstrate that
 if we remove the bound state at $z=z_2$ by using the
 Darboux transformation formulas given in Section~4
 then the transformed potential is no longer compactly supported.
  From
(2.38) we see that the values $\lambda_1$ and $\lambda_2$
corresponding
$z_1$ and $z_2,$ respectively, are given by
$$z_1=-\ds\frac{1}{2}, \quad \lambda_1=\ds\frac{9}{2},\quad z_2=\ds\frac{1}{2}, \quad \lambda_1=-\ds\frac{1}{2}.\tag 5.22$$
Using (5.16)-(5.18) and (5.22) in (4.11) we obtain
$$\theta_n=\left(\ds\frac{1}{2}\right)^{n-1}\bm (-1)^{n-1} \left(5+2\sqrt{5}\right)\\
\stretch
 \left(5-2\sqrt{5}\right)\endbm,\qquad n=1,2,\dots.\tag 5.23$$
Using (5.13) with $N=2$ in (4.10) we obtain
$$F_2=\bm -\ds\frac{9}{10} & 0\\
\stretch 0& -\ds\frac{15}{16} \left(9+4\sqrt{5}\right)\endbm.\tag 5.24$$
With the help of (5.13), (5.23), and (5.24), from (4.17) and (4.18) we can evaluate
the transformed potential $\tilde V_n$ for all $n\ge 1.$ We list the first few
values below and mention that $\tilde V_n$ is not compactly supported:
$$\tilde V_1=\ds\frac{5(3-2\sqrt{5})}{16},\quad
\tilde V_2=\ds\frac{1125+21826\sqrt{5}}{119120},\quad
\tilde V_3=\ds\frac{270(14781+6364\sqrt{5})}{15975481},$$
$$\tilde V_4=\ds\frac{1080(231681+102364\sqrt{5})}{1284143281},\quad
\tilde V_5=\ds\frac{4320(3691281+163364\sqrt{5})}{204372438481}.$$

\vskip 5 pt

\noindent {\bf Acknowledgments.}
The first author expresses his gratitude to the Institute of Physics and Mathematics of the Universidad Michoacana de San Nicol\'as de
Hidalgo, M\'exico for its hospitality.
The second author was partially supported by SNI-CONACYT and CIC-UMNSH, Mexico.

\vskip 5 pt

\noindent {\bf{References}}

\item{[1]} T. Aktosun, P. Sacks, and M. Unlu, {\it Inverse problems for selfadjoint Schr\"odinger operators on the half line with compactly supported potentials,}
J. Math. Phys. {\bf 56}, 022106 (2015).

\item{[2]}
T. Aktosun and V. G. Papanicolaou, {\it Inverse problem with transmission eigenvalues for the discrete Schr\"odinger equation,}
J. Math. Phys. {\bf 56}, 082101 (2015).

\item{[3]}
K. M. Case and M. Kac,
{\it A discrete version of the inverse scattering problem,}
J. Math. Phys. {\bf 14}, 594--603 (1973).

\item{[4]} K. Chadan and P. C. Sabatier, {\it Inverse problems in quantum
scattering theory,} 2nd ed., Springer, New York, 1989.

\item{[5]} R. Courant and D. Hilbert, {\it Methods of mathematical physics,}
Vol. 1, Interscience, New York, 1962.

\item{[6]} D. Damanik and R. Killip, {\it Half-line Schr\"odinger operators with no bound states,} Acta
Math. {\bf 193}, 31--72 (2004).

\item{[7]} D. Damanik and G. Teschl, {\it
Bound states of discrete Schr\"odinger operators with super-critical inverse square potentials,}
Proc. Amer. Math. Soc. {\bf 135}, 1123--1127 (2007).

\item{[8]} G. Darboux, {\it Le\c{c}ons sur la th\'eorie g\'en\'eral des surfaces,} 2nd Part, 2nd ed., Gauthier-Villars, Paris, 1915.

\item{[9]} P. Deift and E. Trubowitz, {\it Inverse scattering
    on the line,} Commun. Pure Appl. Math. {\bf 32}, 121--251
    (1979).

\item{[10]} I. M. Gel'fand and B. M. Levitan, {\it On the
determination of a differential equation from its spectral
function,} Amer. Math. Soc. Transl. {\bf 1} (ser. 2),
253--304 (1955).

\item{[11]}
G. S. Guseinov,
{\it The inverse scattering problem
of scattering theory for a second-order difference equation on the
whole axis,}
Soviet Math. Dokl. {\bf 17}, 1684--1688 (1976).

\item{[12]}
G. S. Guseinov,
{\it The determination of an infinite Jacobi matrix from
the scattering data,}
Soviet Math. Dokl. {\bf 17}, 596--600 (1976).

\item{[13]}
R. Killip and B. Simon,
{\it Sum rules for Jacobi matrices and their applications to spectral theory,}
Ann. of Math. {\bf 158}, 253--321 (2003).

\item{[14]} V. B. Matveev and M. A. Salle, {\it
Darboux transformations and solitons,} Springer-Verlag, Berlin, 1991.

\item{[15]}
V. Spiridonov and A. Zhedanov,
{\it Discrete Darboux transformations,
the discrete-time Toda lattice,
and the Askey-Wilson polynomials,}
Methods Appl. Anal. {\bf 2}, 369--398 (1995).

\item{[16]}
A. A. Suzko,
{\it Darboux transformations for a system of
coupled discrete Schr\"odinger equation,}
Physics of Atomic Nuclei {\bf 65}, 1553--1559 (2002).

\item{[17]} B. N. Zakhariev  and A. A. Suzko, {\it
Direct and inverse problems,} Springer-Verlag, Berlin, 1990.

\end